\documentclass[12pt]{iopart}

\usepackage{iopams}  
\usepackage{amssymb}
\expandafter\let\csname equation*\endcsname\relax
\expandafter\let\csname endequation*\endcsname\relax
\usepackage{amsmath}

\usepackage{cite}
\usepackage{hyperref}
\usepackage{verbatim} 

\usepackage{graphicx}
\newcommand{\kB}{k_{\rm B}}

\newcommand{\ew}{\epsilon_{\rm w}}
\newcommand{\ep}{\epsilon_{\rm p}}
\newcommand{\gw}{\gamma_{\rm w}}
\newcommand{\gp}{\gamma_{\rm p}}
\newcommand{\sumt}{\textstyle \sum}
\newcommand{\cond}{Ddx4$_{\rm cond}$}
\newcommand{\ftoa}{Ddx4$_{\rm 14FtoA}$}

\begin{document}

\title[RPA theory for two charged sequences]{Charge Pattern Matching 
as a ``Fuzzy'' Mode of Molecular Recognition for the Functional Phase
Separations of Intrinsically Disordered Proteins}

\author{Yi-Hsuan Lin}
\address{Department of Biochemistry, 
University of Toronto, and Molecular
Medicine, Hospital for Sick Children, Toronto, Ontario, Canada}
\author{Jacob P. Brady}
\address{Departments of Molecular Genetics, Biochemistry, and Chemistry,
University of Toronto, Toronto, Ontario, Canada}
\author{Julie D. Forman-Kay}
\address{Molecular Medicine, Hospital for Sick Children, 
and Department of Biochemistry, 
University of Toronto, Toronto, Ontario, Canada}
\author{Hue Sun Chan}
\address{Departments of Biochemistry and Molecular Genetics, 
University of Toronto, Toronto, Ontario, Canada}
\ead{chan@arrhenius.med.toronto.edu}
\vspace{10pt}
\begin{indented}
\item[] \today
\end{indented}

\begin{abstract}
Biologically functional liquid-liquid phase separation of intrinsically
disordered proteins (IDPs) is driven by interactions encoded by their amino 
acid sequences. Little is currently known about the molecular recognition 
mechanisms for distributing different IDP sequences into 
various cellular membraneless compartments. Pertinent physics 
was addressed recently by applying 
random-phase-approximation (RPA) polymer theory to electrostatics,
which is a major energetic component governing IDP phase properties. 
RPA accounts for charge patterns and thus has
advantages over Flory-Huggins and Overbeek-Voorn mean-field theories.
To make progress toward deciphering the phase behaviors of multiple IDP 
sequences, the RPA formulation for one IDP species plus solvent is hereby 
extended to treat polyampholyte solutions containing two IDP species. 
The new formulation generally allows for binary 
coexistence of two phases, each containing a different set of volume fractions
$(\phi_1,\phi_2)$ for the two different IDP sequences. 
The asymmetry between the two predicted coexisting phases with 
regard to their $\phi_1/\phi_2$ ratios for the two sequences
increases with increasing mismatch between their charge 
patterns.  This finding points to a  multivalent, stochastic, ``fuzzy'' mode 
of molecular recognition that helps populate various IDP sequences
differentially into separate phase compartments. An intuitive 
illustration of this trend is provided by Flory-Huggins 
models, whereby a hypothetical case of ternary coexistence is also explored. 
Augmentations of the present RPA theory with a relative permittivity 
$\epsilon_{\rm r}(\phi)$ that depends on IDP volume fraction 
$\phi=\phi_1+\phi_2$ 
lead to higher propensities to phase separate, in
line with the case with one IDP species we studied previously. 
Notably, the cooperative, phase-separation-enhancing effects predicted by 
the prescriptions for $\epsilon_{\rm r}(\phi)$ we deem physically plausible 
are much more prominent than that entailed by common effective medium 
approximations based on Maxwell Garnett and Bruggeman mixing formulas. 
Ramifications of our findings on further theoretical development 
for IDP phase separation are discussed.
\end{abstract}

%\maketitle

%%%%%%%%%%%%%%%%%%%%%%%%%%%%%%%%%%%%%%%%%%%%%%%%%%%%%%%%%%%%%%%%%%%%%%%%%%%%%%%%

\section{Introduction}

Nearly two decades of increasingly intensive research established
that intrinsically disordered proteins/protein
regions (abbreviated collectively as IDPs here) serve many important
biological functions, and are especially critical for signaling and regulation 
in multicelluar organisms \cite{Uversky00,Dunker01,Tompa02,Nussinov2003,tompa12,vanderLee14,Liu2014,Chen15,PappuCOSB,Wright2015,Veronika2016,alaji2016}.
Recently, it was discovered that IDPs function not only at the level of 
individual molecules. An expanding repertoire of IDPs have been seen to undergo 
liquid-liquid phase separation in vitro, intriguingly parallelling the 
formation of many types of condensed liquid/gel-like bodies/organizations in 
living organisms, including extracellular materials, transcription complexes,
nucleating sites of intermediate filament organization,
and various membraneless organelles. It is apparent from
these experimental observations that IDP condensation constitutes a major 
physical underpinning of these condensed bodies, which serve as 
hubs for specifically regulated sets of biomolecules to interact. 
As such, IDP phase separation is one of
Nature's means to achieve the spatial and temporal compartmentalization 
necessary for the organization of vital processes \cite{Brangwynne2009,Rosen12,McKnight12,Lee13,Hyman14,wright14,Seydoux2014,Nott15,ElbaumGarfinkle15,tanja2015,Alberti2015,Nott16,Rosen2016,RosenPappu2016,Feric16,Mitrea16,Fuxreiter2016,Michnick2016,Riback_etal2017,cliff2017}.
Although much detail remains to be ascertained,
examples of membraneless organelles and an IDP species whose
phase separation has been found to be a likely contributor to 
their assembly include chromatoid bodies, nuage or 
germ granules in mammalian male germ cells \cite{Kotaja2011}
and the DEAD-box RNA helicase Ddx4 \cite{Nott15}, 
the {\it Caenorhabditis elegans} germline-specific perinuclear
RNA granules known as P-granules \cite{Wang2014}
and the Ddx3 RNA helicase LAF-1 \cite{ElbaumGarfinkle15},
as well as the stress granules triggered by integrated stress 
response \cite{anderson2013} and the RNA-binding heterogeneous 
nuclear ribonucleoprotein A1 (hnRNPA1) \cite{tanja2015}.
Because of the importance of these bodies to biological regulation,
malfunctioning of the corresponding IDP phase separation 
processes can lead to deregulation and diseases, including
cancer due to loss of regulation of stress 
granules~\cite{ivanov2015}, protein 
fibrillization and thus amyloid diseases \cite{tanja2015},
and various forms of neurological disorder \cite{Li2013,Parker2013,mingjie2016}.

\subsection{Seeking ``sequence-phase'' relationships}

With the advent of IDPs, the molecular biology paradigm of seeking 
``sequence-structure'' relationships for globular proteins has to be
expanded to encompass ``sequence-ensemble'' relationships for individual 
IDPs \cite{TanjaJulie2013,Neale2013,cider2017}. 
Now, the additional question we need to ask is: How do
the phase behaviors of IDPs depend on their amino acid sequences? In other 
words, what are the ``sequence-phase'' relationships \cite{alex15}? 
Although computational study of IDPs is still in its infancy, 
much insight into the conformational properties and binding energetics of 
individual IDPs has been gained by explicit-chain simulations
\cite{JChen2012,Song13,Best2014,Chen15,PappuCOSB,jianhui15,Best2017,Shea2017}.
In contrast, because IDP phase separation is a multiple-chain 
property, computationally it is extremely costly to simulate using
fully atomic explicit-chain models~\cite{Regis2006}, notwithstanding
promising progress made by coarse-grained approaches that treat groups 
of amino acid residues of IDPs as interaction modules in continuum space
\cite{Ruff15} or on lattices \cite{Feric16} and 
simulation algorithms developed recently \cite{HX_zhou16,HX_zhou17} 
for phase separations of globular proteins \cite{vojko15}.
In this context, analytical theories of IDP phase separation are valuable not
only because of their tractability, but also---and more importantly---for
the conceptual framework they offer for understanding a highly complex 
phenomenon.  

\subsection{Mean-field and random-phase-approximation (RPA) theories of
phase coexistence}

Developed mainly for synthetic polymers at its inception, the basic
statistical mechanical framework of Flory-Huggins (FH) 
theory \cite{Flory53,Chan94} 
is useful for describing phase separation in the biomolecular 
context \cite{Rohit08,Hyman14,Nott15,Brangwynne15}. FH assumes that
the interactions among the monomers (residues) of the polymers have
short spatial ranges. In view of the fact that not only short-range
interactions such as solvent-mediated hydrophobic effects but 
long-range Coulomb forces 
are important for driving the phase separation of certain IDPs, it has been 
suggested \cite{Brangwynne15} that the Overbeek-Voorn (OV) 
theory \cite{Voorn56,OverbeekVoorn57} should be more appropriate in 
those cases. Inasmuch as sequence dependence is concerned, however,
both FH and OV are mean-field theories that account only for composition 
but not sequence information. In these theories, all residues 
belonging to any given set of chain sequences are allowed to interact on equal 
footing irrespective of the correlation dictated by chain connectivity.
Therefore, to address sequence specificity of IDP phase separation, one
needs to go beyond FH and OV. Accordingly, we recently put forth a 
random-phase approximation (RPA) theory that approximately accounts 
for the effects of arraying different charge patterns along the IDP chain 
sequence \cite{Lin16,Lin17a,Lin17b}. The term ``RPA'' was first introduced by
Bohm and Pines in their quantum mechanical collective
description of electron interactions \cite{BohmPines1951}. The analogy
of this approach with the approximate polymer theory that considers 
terms up to quadratic in particle density was recognized by
de Gennes\cite{deGennes79}. As detailed elsewhere \cite{Lin17a}, 
the RPA formulation we developed \cite{Lin16}, which follows largely that 
of Olvera de la Cruz\cite{Mahdi00,Ermoshkin03}, 
is successful in providing a physical 
rationalization \cite{Lin16,Lin17a} for the experimental salt 
dependence of Ddx4 phase separation as well as the difference in phase 
behavior between the wildtype and a charge-scrambled variant of 
Ddx4 \cite{Nott15}. Our RPA theory suggests further that the tendency for a 
collection of IDP chains with a given charge sequence to phase separate is 
strongly---but negatively---correlated with the conformational 
dimensions of individual IDP molecules of the sequence \cite{Lin17b},
and that both of these properties are well correlated with the 
charge pattern parameter $\kappa$ of Das and Pappu \cite{Das13}
and the ``sequence charge decoration'' parameter SCD of Sawle and 
Ghosh \cite{Sawle15}. In principle, these predictions are now testable 
using experimental techniques similar to those 
employed to study ``IDP polymers'' \cite{chilkoti2015}.

\subsection{Sequence-dependent multiple-component IDP phase separation}

Membraneless organelles are complex functional units
consisting of many protein and nucleic acid components \cite{wright14}. 
Different types of such units are enriched with different varieties of 
proteins and nucleic acids. Some individual membraneless organelles 
have mesoscopic substructures with different degrees of fluidity, as in 
the case of stress granules \cite{parker2016}. In a similar vein, immiscible 
liquid phases have been shown to contribute to subcompartmentalization 
of the nucleolus \cite{Feric16}. Clearly, a viable spatial organization
of cellular processes necessitates a heterogeneous distribution of different 
biomolecular components into different membraneless organelles and 
their substructures, rather than having all IDP species condensing
into a big {\it gemisch}. How is this achieved physically?

One aspect of this question was addressed recently using a simple cubic
lattice model of multicomponent mixtures confined to a $6\times 6\times 6$ 
box, with each component represented by a bead on the 
lattice \cite{frenkel2017}. By considering hypothetical intercomponent 
interaction strengths (contact energies), the authors found that with 
sufficient heterogeneity in contact energies, demixed domains are likely 
to segregate. This and other results of this big-picture study suggest that
phase separation into cellular compartments with different compositions
is a robust consequence of interaction heterogeneity among
biomolecules \cite{frenkel2017,alex17}.

With this in mind, the logical next step in our pursuit of sequence-phase
relationships is to ascertain how interaction heterogeneity is encoded
genetically. For globular protein folding, the fact that the amino acid 
alphabet is finite \cite{chan1999} implies that there are physical limits to 
interaction heterogeneity and structural encodability, as has been
illustrated by simple exact models \cite{chandill1996,Wroe05}. Similarly,
physical limits should exist in the ability of different IDP sequences 
to demix. Taking a step toward deciphering what is physically achievable,
here we present an RPA formulation for the phase behavior of two 
charged sequences as models for two IDP species. 
Consistent with physical intuition, we found that the tendency for the 
two IDP species to demix in two coexisting phases increases with increasing 
difference in their charge patterns. This phenomenon represents a statistical, 
multivalent mode of molecular recognition for cellular organization 
that differ from the structurally highly specific form of recognition 
among folded proteins but share similarities with the 
``fuzzy complexes'' \cite{fuzzy08,fuzzy15,fuzzy17}
involving individual IDP molecules \cite{borg07,tanja2010,larry14,Veronika2017}.

Delving deeper into the role of electrostatics in IDP phase separation,
we have also extended the two-sequence RPA formulation to address how a 
relative permittivity $\epsilon_{\rm r}(\phi)$ that depends on IDP 
volume fraction $\phi$ may affect IDP phase properties. Several common 
effective medium approximations \cite{Markel16} posit a gradual decrease in 
$\epsilon_{\rm r}$ from the $\epsilon_{\rm r}(\phi=0)\approx 80$ value
for pure water with increasing $\phi$. But physical consideration \cite{Lin17a} 
and experimental volumetric measurements suggest a much sharper decrease, 
with $\epsilon_{\rm r}(\phi=0.2)\approx 20$, leading to large 
cooperative effects that enhance phase separation significantly. 
These findings and their ramifications are detailed below.

%%%%%%%%%%%%%%%%%%%%%%%%%%%%%%%%%%%%%%%%%%%%%%%%%%%%%%%%%%%%%%%%%%%%%%%%%%%%%%%%

\section{Methods}
\label{sec:methods}

\subsection{Theoretical development of the RPA formulation}

The development of RPA theory for a pair of charged sequences
constitutes the bulk of the results presented in subsequent sections 
of this article. This effort is based on an extension of the RPA 
formulation for a single sequence that we put forth 
recently \cite{Lin16,Lin17a}.

\subsection{Experimental determination of dissolved protein volumes}

To address the effect of volume of dissolved proteins on the relative 
permittivity of the resulting aqueous solution, nuclear magnetic
resonance (NMR) and 
absorbance measurements were performed on two folded 
globular proteins bovine serum albumin (BSA) and hen egg white lysozyme 
(HEWL) as well as two IDPs \ftoa and \cond, which are, respectively, a 
mutant of Ddx4 in which all 14 phenylalanines are mutated to alanines and
the concentrated phase of phase-separated wildtype Ddx4.

\subsection{Measurement of water content of protein samples}

NMR spectra were recorded using a Bruker 
Ascend III spectrometer at 14.0 T equipped with a cryogenically cooled 
triple resonance gradient probe.
Spectra were processed using NMRPipe \cite{Delaglio:1995aa}.
1D $^1$H spectra were recorded on protein samples over a range of 
concentrations (5-400 mg mL$^{-1}$) and the integrated water signals 
were compared with the corresponding integrals obtained from a spectrum 
recorded of buffer (the same buffer composition as used for each 
protein sample).
BSA was purchased from Sigma and samples were 
prepared in 20 mM sodium phosphate (NaPi), 100 mM sodium chloride (NaCl), 
10 \% $^2$H$_2$O/90 \% $^1$H$_2$O, pH 6.5.
HEWL was purchased from BioBasic and samples 
were dissolved in 20 mM sodium citrate, 100 mM NaCl, 
10 \% $^2$H$_2$O/90 \% $^1$H$_2$O, pH 5 (lower pH was used due to 
limited solubility of HEWL in NaPi at pH 6.5).
\ftoa samples were prepared according to \cite{Brady:2017} and dialysed 
against 20 mM NaPi, 100 mM NaCl, 5 mM tris(2-carboxyethyl)phosphine (TCEP), 
10 \% $^2$H$_2$O/90 \% $^1$H$_2$O, pH 6.5. For \cond the same buffer 
was used but the NaCl concentration  was varied between 100-400 mM in 
order to generate samples with protein concentrations between 200 
and 400 mg mL$^{-1}$.
For phase-separated samples, it was ensured that the entirety of the 
probe coil was occupied by the condensed phase, thus avoiding 
contaminating signals from the more hydrated dilute phase.
Spectra were recorded using both small flip angle 
($\theta$ $<$ 10$^{\circ}$) and $\theta$ = 90$^{\circ}$ pulses with 
very similar results in both cases.

\subsection{Measurement of protein concentration}
Protein concentrations were determined by absorbance at 280 nm (A$_{280}$) 
after dilution into 6 M guanidinium HCl, 20 mM NaPi, pH 6.5 using the 
Beer-Lambert law with extinction coefficients of 23 950, 23 950, 36 000, and 
44 309 M$^{-1}$cm$^{-1}$ for wildtype Ddx4, \ftoa, HEWL, and BSA, 
respectively \cite{Gill:1989aa}.

%%%%%%%%%%%%%%%%%%%%%%%%%%%%%%%%%%%%%%%%%%%%%%%%%%%%%%%%%%%%%%%%%%%%%%%%%%%%%%%%

\section{Overview of three-component phase behaviors}

Possible phase behaviors of a three-component liquid system are
outlined in Fig.~1. In general, the system can be a homogeneous 
solution [Fig.~1(b], or it can separate into two coexisting phases [binary 
coexistence; Fig.~1(c)--(e)], or separate into three coexisting 
phases [ternary coexistence; Fig.~1(f)]. Fundamentally, phase behavior
is governed by the intra- and inter-component interactions as well 
as environmental conditions such as temperature and pressure.
Our theories below provide a rudimentary physical account of 
how interactions among IDP chains with two different amino acid sequences
affect the conditions under which binary and ternary coexistence emerge.
The theories presented here are for solution systems
with an effective infinite volume. As such, our theories 
account for the differences among scenarios typified by the leftmost 
drawings in Fig.~1(c)--(f) but they are not equipped to address details 
such as droplet size and geometry. In other words, 
they provide no discrimination among different droplet geometries 
along a given horizontal row in Fig.~1. Accounting for the latter would 
require additional modeling of the interfacial tensions between different 
solution phases \cite{Feric16}. 

\begin{figure}%[htbp]
\includegraphics[width=0.83\columnwidth]{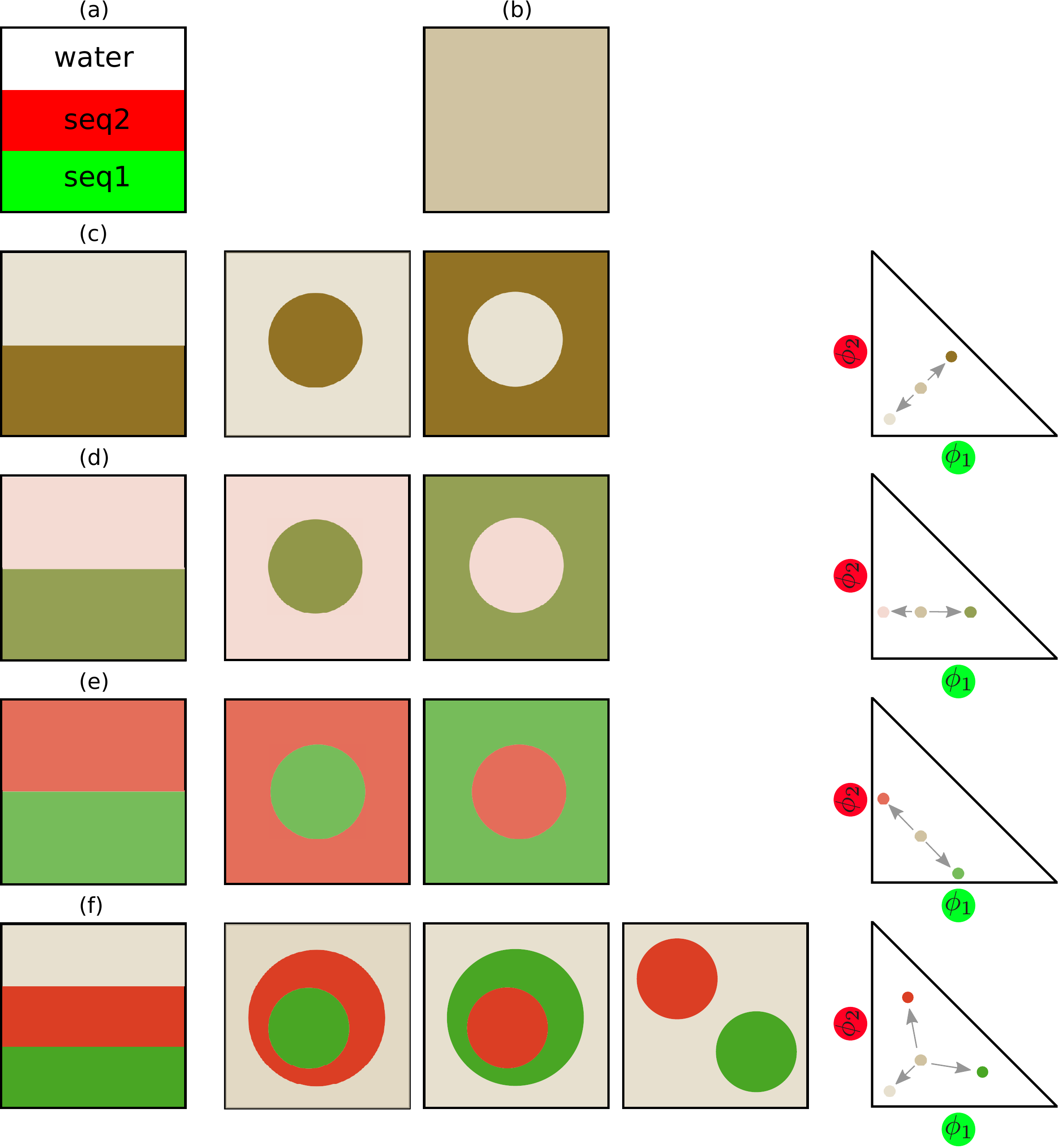}
\caption{Schematics of three-component phase separation scenarios. (a) 
The three pure components considered in our analysis---water solvent, 
IDP sequence 1 (seq1), and IDP sequence 2 (seq2)---in a totally demixed 
configuration.
(b) A homogeneous solution of all three components. In this and
subsequent drawings in this figure, the colors of various solution
phases are resultant colors of mixing the pure components (white, red, and
green) in proportions consistent with the given scenario.
(c)--(e) Binary coexistence. (c) One of the coexisting phases is dilute
in both seq1 and seq2 (top), the other is concentrated in both seq1 and
seq2 (bottom). (d) The concentration of seq2 is constant throughout,
whereas seq1 is dilute in one phase (top) and concentrated in the
other (bottom). (e) One phase is concentrated in seq2 but dilute
in seq1 (top), the other phase is concentrated in seq1 but dilute
in seq2 (bottom). (f) Ternary coexistence. In this example, the
system separates into a dilute phase for both seq1 and seq2 (top),
a phase concentrated in seq2 but dilute in seq1 (middle), and
a phase concentrated in seq1 but dilute in seq2 (bottom), with water 
present in all three phases. Circle(s)-in-square drawings to the
right of (c)--(f) depict possible configurations of phase-separated
droplets in the given scenario. In each case, the rightmost schematic 
phase diagram illustrates the manner in which a given set of bulk 
concentrations (bulk volume fractions for seq1 and seq2) of 
$(\phi_1,\phi_2)$ represented by the central dot with the color of (b) 
is separated (arrows) into two or three phases with different
$(\phi_1,\phi_2)$'s. The volume fraction of water is equal to
$1-\phi_1-\phi_2$.
}
   \label{fig:drawingRR}
\end{figure}

\section{RPA theory for two charged sequences}

Based on our previous RPA formulation for a single 
sequence \cite{Lin16,Lin17a}, the approach is now extended to consider 
two model polypeptide (IDP) sequences $s_1$ and $s_2$ in a salt-free aqueous 
solution. Using notation similar to before \cite{Lin17a}, the electric 
charges along the sequences are written as 
$\{ \sigma_1\} \equiv \{ \sigma_1^{(1)}, \sigma_1^{(2)}, \sigma_1^{(3)}, 
\dots, \sigma_1^{(N_1)} \}$ and 
$\{ \sigma_2 \} \equiv \{ \sigma_2^{(1)}, \sigma_2^{(2)}, \sigma_2^{(3)}, 
\dots, \sigma_2^{(N_2)} \}$,
where $N_1$ and $N_2$ are the numbers of residues in $s_1$ and $s_2$,
respectively. The corresponding volume fractions of the IDPs in solution are
denoted as $\phi_1$ and $\phi_2$. The present study is restricted to
IDP sequences with zero net charge. Counterions are not considered.

\subsection{Free energy as a function of two IDP volume fractions}

Following the FH lattice argument~\cite{Flory53}, we partition the 
spatial volume $V$ of the solution system into lattice units, $a^3$, that
corresponds to the volume of a solvent molecule. Accordingly, the 
RPA free energy per unit volume and in units of the product of
Boltzmann constant $k_{\rm B}$ and absolute temperature $T$ is cast 
as the per-lattice-site quantity
\begin{equation}
f(\phi_1,\phi_2) \equiv \frac{F_{\rm RPA}a^3}{V \kB T} 
= -s(\phi_1,\phi_2)  + f_{\rm el}(\phi_1,\phi_2) ,
\label{eq1}
\end{equation}
where the negative entropy $-s$ is the entropic contribution to 
free energy in units of $k_{\rm B}T$. This term is given by the standard 
FH entropy of mixing for a system comprising of $s_1$, $s_2$, and solvent:
\begin{equation}
-s(\phi_1,\phi_2)  = \frac{\phi_1}{N_1}\ln \phi_1 + 
\frac{\phi_2}{N_2}\ln \phi_2 + (1-\phi_1-\phi_2)\ln(1-\phi_1-\phi_2) \; ,
\end{equation}
where $1-\phi_1-\phi_2$ is the volume fraction of solvent.
The electrostatic contribution $f_{\rm el}$ is calculated by 
RPA~\cite{Lin16, Lin17a}, viz.,
\begin{equation}
f_{\rm el}(\phi_1,\phi_2)  = \int_0^\infty \frac{d k k^2}{4\pi^2} \left\{ \ln \left[ 1+ {\cal G}(k) \right] - {\cal G}(k) \right\},
	\label{eq:fel_integral}
\end{equation}
where $k$ is the reduced wave number that absorbs the
virtual bond length $b\simeq a$ of the polypeptide backbone by re-defining
$kb=\tilde k$ in Ref.~\cite{Lin17a} as $k$ (i.e., $kb \to k$), 
such that
\begin{equation}
{\cal G}(k) = \frac{4\pi}{k^2(1+k^2)T^*}\langle q | \hat{G}_k | q \rangle,
	\label{eq:Gbracket}
\end{equation}
where $4\pi/[k^2(1+k^2)]$ is from the Fourier transformation of 
Coulomb interaction with a short-range cutoff~\cite{Ermoshkin03, Lin16}
in units of $\kB T$,
\begin{equation}
U_{\rm el}(r) = \frac{e^2}{4\pi\epsilon_0\epsilon_{\rm r}\kB T}\frac{1-e^{-r/b}}{r} \; ,
\end{equation}
$\epsilon_0$ is vacuum permittivity and $\epsilon_{\rm r}$ is relative 
permittivity, $T^* \equiv b/l_B$ is the reduced temperature defined by 
Bjerrum length 
$l_B = e^2/(4\pi\epsilon_0\epsilon_{\rm r} \kB T)$. Here $| q \rangle$ is the 
$(N_1+N_2)$-dimensional column vector representing the two charge 
sequences, namely $q_i = \sigma_1^{(i)}$ for $1\leq i\leq N_1$ 
and $q_i = \sigma_2^{(i-N_1)}$ for $N_1\!+\!1 \leq i\leq N_1\!+\!N_2$, 
$\langle q |$ is the transposed row vector, 
$\langle q | \hat{G}_k |q\rangle \equiv \sum_{ij} q_i(\hat{G}_k)_{ij}q_j$ 
with $(\hat{G}_k)_{ij}$ 
being the $i,j$ element of the {\em bare} two-body correlation matrix
$\hat{G}_k$ of all possible sequence-sequence correlations~\cite{Lin17a},
\begin{equation}
\hat{G}_k = \left(
\begin{array}{cc}
\hat{G}_{11}(k) & \hat{G}_{12}(k), \\
\hat{G}_{21}(k) & \hat{G}_{22}(k)
\end{array}
\right),
\end{equation}
and $\hat{G}_{12}(k) = \hat{G}_{21}(k)$. 
As in our previous studies \cite{Lin16,Lin17a}, we consider a simple 
formulation in which all IDPs are modeled as Gaussian chains without 
excluded volume within the RPA formalism~\cite{Mahdi00}. In this
approximation, there is no correlation between the positions of different 
chains; hence $\hat{G}_{12}(k) = \hat{G}_{21}(k) =0$ and
\begin{equation}
\begin{aligned}
\hat{G}_{11}(k)_{ij} = & \frac{\phi_1}{N_1}\exp\left( -\frac{1}{6}k^2|i-j|\right) \\
\hat{G}_{22}(k)_{ij} = & \frac{\phi_2}{N_2}\exp\left( -\frac{1}{6}k^2|i-j|\right)
\end{aligned}
\end{equation}
follow from the average of $\exp(i{\bf k}\cdot{\bf R}_{ij})$
over a Gaussian chain ensemble wherein
${\bf R}_{ij}$ is the vector between chain positions $i,j$ and 
$k^2={\bf k}\cdot{\bf k}$ (Eq.~(IX.59) of \cite{deGennes79}).
Eq.~(\ref{eq:Gbracket}) then becomes
\begin{equation}
{\cal G}(k) = \frac{4\pi}{k^2(1+k^2)T^*}
\left[ 
\langle \sigma_1 | \hat{G}_{11}(k) | \sigma_1 \rangle  + 
\langle \sigma_2 | \hat{G}_{22}(k) | \sigma_2 \rangle 
\right] \; ,
\label{calG_def}
\end{equation}
where
\begin{subequations}
\begin{align}
\langle \sigma_1 | \hat{G}_{11}(k) | \sigma_1 \rangle
	 = & \frac{\phi_1}{N_1}
	\sum_{i,j=1}^{N_1} \sigma_1^{(i)}  
\sigma_1^{(j)} \exp\left( -\frac{1}{6}k^2|i-j|\right) \; ,
\label{G_eqsa} \\
\langle \sigma_2 | \hat{G}_{22}(k) | \sigma_2 \rangle
	= & \frac{\phi_2}{N_2} 
	\sum_{i,j=1}^{N_2} \sigma_2^{(i)}  \sigma_2^{(j)} \exp\left( -\frac{1}{6}k^2|i-j|\right) \; .
\label{G_eqsb}
\end{align}
\end{subequations}

\subsection{Free energy landscape and spinodal instability}

\begin{figure}%[htbp] 
\includegraphics[width=\columnwidth]{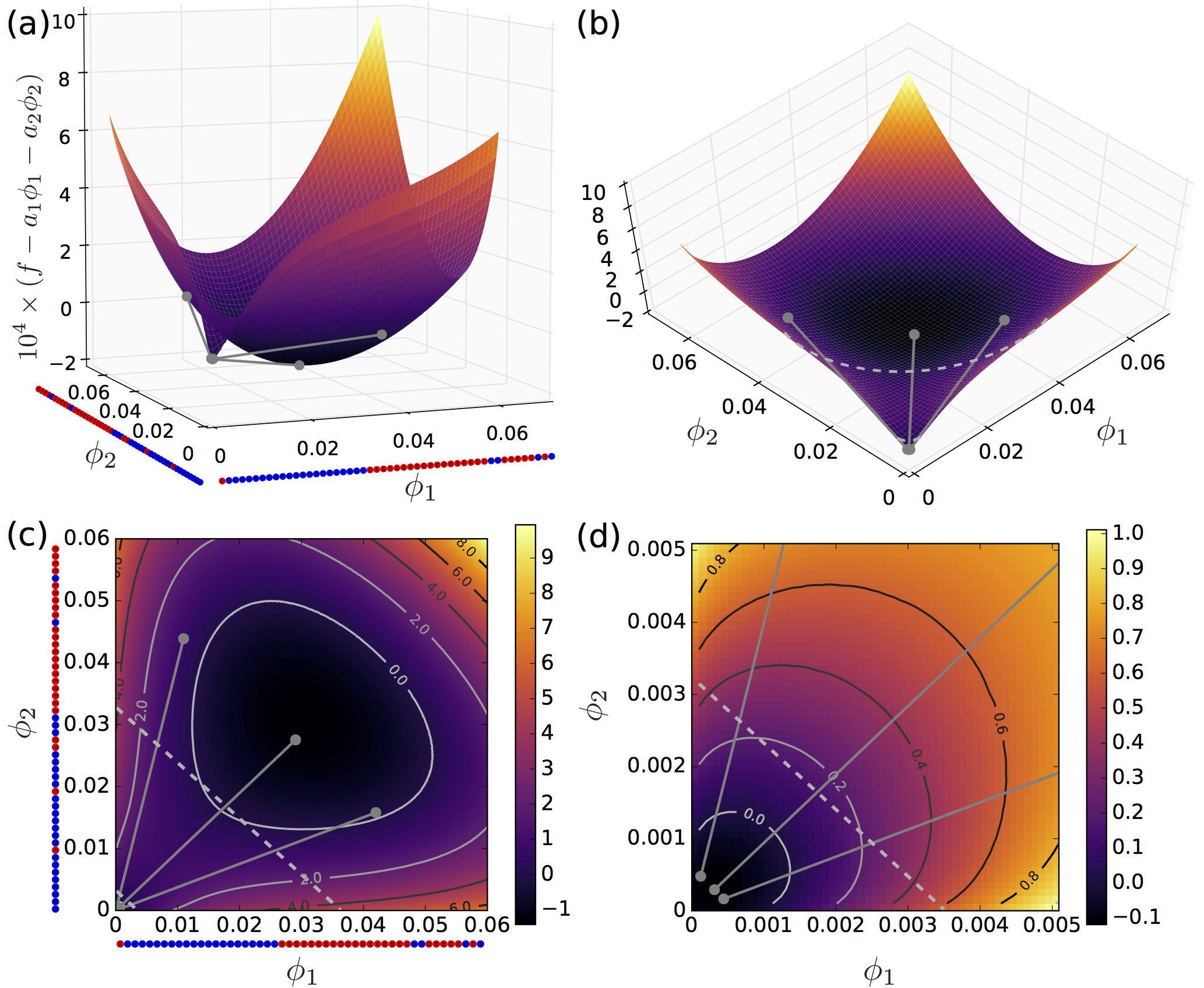} 
\caption{RPA theory of phase separation for a pair of sequences.
The sequences, shown along the axes in (a) and (c), wherein positive
and negative charges are depicted as red and blue beads respectively,
correspond to sv28 and sv24 in Das and Pappu \cite{Das13}. 
$\phi_1$ and $\phi_2$ are their volume fractions, respectively, 
as the sequences are re-labeled as seq1 and seq2 in this work
(Table~\ref{table:SCDofSeqs}). Results in this figure are for $T^*=4$. 
(a) Free energy landscape. The plotted quantity is the free energy
$f(\phi_1,\phi_2)$ in Eq.~(\ref{eq1}) minus a linear function 
$a_1 \phi_1 + a_2 \phi_2$ of $\phi_1$, $\phi_2$ where the coefficients
are chosen to be $a_1 = -1.1447$ and $a_2=-1.1453$. The sole purpose
of subtracting this linear term is to graphically highlight the changes 
in curvature on the landscape. The subtraction has no effect on the
determination of phase coexistence (Sec.~\ref{sec:Binary_3comp}).  
As examples of phase coexistence, three pairs of phase-separated states 
are marked by grey dots, whereby each pair is connected by its own tie line 
(grey solid line). 
(b) View of the landscape in (a) from an elevated vantage. The 
two dashed curves are spinodal phase boundaries defined by 
Eq.~(\ref{eq:spinodal_cond}) (one of the boundaries
is very close to the origin with $\phi_1,\phi_2$ intercepts 
$\approx 0.003$, see zoom-in view in (d)); 
the region between these two boundaries
satisfies $\det\hat{\cal F} < 0$. The three pairs of dots marking 
coexisting phases connected by tie lines are the same as those in (a). 
Note that subtracting a linear function of $\phi$'s from $f(\phi_1,\phi_2)$ 
does not alter the spinodal boundaries determined by 
the matrix $\hat{\cal F}$ that contains only second derivatives of $\phi$'s. 
(c) Heat map representation (scale on the right) of the function 
$10^4 \times [ f(\phi_1,\phi_2)-a_1 \phi_1 - a_2 \phi_2 ]$ in (a).
Pairs of coexisting $(\phi_1,\phi_2)$ connected by tie lines and 
spinodal boundaries are the same as those in (b).
(d) Zoom-in plot of the $0\leq\phi_1,\phi_2\leq0.005$
region in (c) to provide a clearer view of 
the spinodal phase boundary that is very near to the origin
(dashed line) and the dilute phases of the three coexisting pairs.
}
   \label{fig:f_2D3D}
\end{figure}

As an example, we first apply this formulation to two $N_1=N_2=50$ 
charged IDP sequences corresponding to sv28 and sv24 in 
Das and Pappu \cite{Das13}. The sequences are labeled here as seq1 and seq2
respectively.
The RPA free energy function for the two sequences (Fig.~\ref{fig:f_2D3D}) 
consists of regions of different curvatures: parts of the landscape
are convex downward (i.e., has a convex downward curvature) whereas some 
other parts are convex upward (concave downward).

If a given IDP solution has overall (bulk-state) IDP volume fractions 
$(\phi_1^0, \phi_2^0)$ situated in a convex-downward region, the state is 
thermodynamically stable and thus the bulk-state volume fractions are
maintained. In contrast, if the bulk state is in a concave-downward 
region, the state is thermodynamically unstable because
a phase-separated state allows for a lower free energy. Accordingly,
the system undergoes phase separation to multiple coexisting phases 
with different protein volume fractions~\cite{Lin17a}.

The boundary between the convex and concave regions is determined by 
the saddle point condition \cite{Voorn56c,Lin17a} 
\begin{equation}
\det \hat{\cal F} \equiv
\begin{vmatrix}
\mbox{\large $\frac{\partial^{2}f}{\partial \phi_1^2}$}  & 
\mbox{\large $\frac{\partial^{2}f}{\partial \phi_1 \partial \phi_2}\; $}\\
&\\
\mbox{\large $\; \frac{\partial^{2}f}{\partial \phi_2 \partial \phi_1}$} & 
\mbox{\large $\frac{\partial^{2}f}{\partial \phi_2^2}$}
\end{vmatrix} = 0 \; .
	\label{eq:spinodal_cond}
\end{equation}
This boundary defines a {\em spinodal} instability region in 
which $\det \hat{\cal F} < 0$. Because the determinant of a matrix 
is equal to the product of all its eigenvalues, this instability 
condition indicates that one, but not both, of the eigenvalues 
of $\hat{\cal F}$ is negative. This means that second-order perturbations 
of free energy with respect to $\phi_1,\phi_2$ along the direction of 
the corresponding eigenvector diverges, signaling that the system 
cannot maintain a homogeneous phase. 

Note that although the opposite of the $\det \hat{\cal F} < 0$
condition, viz., $\det \hat{\cal F} > 0$, does not by itself
exclude the possibility that both eigenvalues of $\hat{\cal F}$ are
negative and thus the system is thermodynamically unstable despite
not satisfying $\det \hat{\cal F} < 0$ (e.g. at a local maximum), 
exhaustive numerical searches
($\phi_1,\phi_2=0.001,0.002,\dots,0.999$) did
not find any such instance for all RPA and FH systems studied in
this work. 
Hence, for these systems, Eq.~(\ref{eq:spinodal_cond}) is the 
valid condition for spinodal boundaries, examples of which are shown as 
dashed curves in Fig.~\ref{fig:f_2D3D} for the (seq1, seq2) system.

In two-component systems such as those consisting of a single IDP species
and solvent molecules, spinodal instability necessarily leads to 
binary coexistence~\cite{Lin17a}. In systems 
with more than two components, more coexisting phases are allowed.
According to the Gibbs phase rule, the maximum number of 
coexisting phases under 
a given set of environmental conditions is equal to the number 
of components in the system ~\cite{Lin17a,Atkins09} but,
depending on system specifics, the
actual number of coexisting phases can be smaller. 
Spinodal instability always implies that 
the system existing as a single phase is thermodynamically untenable 
and thus phase separation must occur. However,
when the number of components is larger than two, 
the precise number of 
coexisting
phases is governed by how total free energy varies
with changes in the component volume fractions in different 
combination of phases or, equivalently, by the balance of chemical 
potentials for each component across different phases~\cite{Lin17a, Voorn56c}. 
As examples, three instances of binary coexistence in the 
(seq1, seq2) system are depicted in Fig.~\ref{fig:f_2D3D} as pairs of
dots connected by solid tie lines. 
The general procedure for determining the conditions for such 
coexistence is as follows.

\subsection{Binary coexistence in three-component systems: Applications to 
IDPs with two different sequences plus solvent}\label{sec:Binary_3comp}

As mentioned, we use two methods to determine phase equilibrium:
by ascertaining the minimum free energy among single- and 
multiple-phase states~\cite{Kastelic16}, and 
by balancing the chemical potentials for each of the components across
different phases~\cite{Chan94, Voorn56, Grosberg94}. The two approaches 
are mathematically equivalent. They can be applied simultaneously to
yield more accurate numerical results. Here we describe in detail how 
the two approaches are applied to study three-component systems that 
undergo binary phase separation.

A system is dictated by thermodynamics to seek its lowest free energy state.
Whether a system phase separates can be ascertained by
comparing its overall free energy with and without phase separation. For a
system with bulk IDP volume fractions $(\phi_1^0, \phi_2^0)$, the
free energy $f_{\rm bulk}$ without phase separation and the
free energy $f_{\rm sep}$ for separating into two phases (labeled as
$\alpha$ and $\beta$) that take up fractional volumes $v_\alpha=v$
($0\le v\le 1$) and $v_\beta=1-v$ of the total system volume and with 
IDP volume fractions $(\phi_1^\alpha,\phi_2^\alpha)$ and 
$(\phi_1^\beta,\phi_2^\beta)$, respectively, are given by 
\begin{subequations}
\begin{align}
& f_{\rm bulk} = f(\phi_1^0,\phi_2^0), \\
& f_{\rm sep} = v f(\phi_1^\alpha,\phi_2^\alpha) + 
(1-v) f(\phi_1^\beta,\phi_2^\beta), 
	\label{eq:f_2sep}
\end{align}
	\label{eq:f_bulk_sep}%
\end{subequations}
wherein conservation of volume of each of the components implies that 
\begin{subequations}
\begin{align}
v \phi_1^\alpha + (1-v)\phi_1^\beta & = \phi_1^0 \; , \\
v \phi_2^\alpha + (1-v)\phi_2^\beta & = \phi_2^0 \; .
\end{align}
	\label{eq:density_conserve}%
\end{subequations}
By rewriting Eq.~(\ref{eq:density_conserve}) to express IDP volume
fractions in $\beta$ as functions of $v$ with $0<v<1$
and volume fractions in $\alpha$:
\begin{subequations}
\begin{align}
\phi_1^\beta & = \frac{\phi_1^0 - v \phi_1^\alpha}{1-v}, 
\label{eq:phi_1_beta}\\ 
\phi_2^\beta & = \frac{\phi_2^0 - v \phi_2^\alpha}{1-v}, 
\label{eq:phi_2_beta}
\end{align}
\end{subequations}
$f_{\rm sep}$ is seen as a function
of $\phi_1^\alpha$, $\phi_2^\alpha$, and $v$,
\begin{equation}
f_{\rm sep}(v,\phi_1^\alpha, \phi_2^\alpha) =
	 v f(\phi_1^\alpha,\phi_2^\alpha) 
	 + (1-v) f\left(\frac{\phi_1^0- v \phi_1^\alpha}{1-v},
\frac{\phi_2^0 - v \phi_2^\alpha}{1-v}\right) \; .
	 \label{eq:f_sep_3var}
\end{equation} 
Note that $\lim_{v\to 0}
f_{\rm sep}(v,\phi_1^\alpha, \phi_2^\alpha)
=\lim_{v\to 1}f_{\rm sep}(v,\phi_1^\alpha, \phi_2^\alpha)= 
f(\phi_1^0,\phi_2^0)$, where 
$(\phi_1^\beta,\phi_2^\beta)\to (\phi_1^0,\phi_2^0)$ for $v\to 0$ and
$(\phi_1^\alpha,\phi_2^\alpha)\to (\phi_1^0,\phi_2^0)$ for $v\to 1$.

To find the set of variables that yields the global minimum of $f_{\rm sep}$, 
we numerically search the three-dimensional space of 
$(v,\phi_1^\alpha,\phi_2^\alpha)$ by implementing the sequential least
squares programming (SLSQP) algorithm~\cite{Kraft88} using the {\tt
scipy.optimize.minimize} function in Scipy, a Python-based numerical package
for scientific computation~\cite{Scipy}. If a given 
$(v,\phi_1^\alpha,\phi_2^\alpha)$ is found 
to yield a minimum of $f_{\rm sep}$ among computed $f_{\rm sep}$ values
and also satisfies 
$f_{\rm sep}(v, \phi_1^\alpha,\phi_2^\alpha) < f_{\rm bulk}(\phi_1^0,
\phi_2^0)$, the system is judged to be in a state
of binary phase separation to the two phases $\alpha$ and $\beta$. In 
contrast, if all $f_{\rm sep}$ for a given $(\phi_1^0, \phi_2^0)$ are 
larger than $f_{\rm bulk}$, the bulk state $(\phi_1^0, \phi_2^0)$ is 
thermodynamically stable and the system does not phase separate. Unlike in the
two-component case in which the two separated phases are unique and
independent of the bulk IDP concentration/volume fraction insofar as it is 
in the phase-separated regime, in three-component systems different bulk 
IDP concentrations/volume fractions can result in different $\alpha, \beta$ 
phases. Thus a complete binary phase diagram is generated by considering 
all possible $(\phi_1^0, \phi_2^0)$ combinations.

For two-component systems (one IDP sequence plus solvent), we have shown that
linear terms of $\phi$ in the system free energy do not affect the
determination of phase equilibrium~\cite{Lin17a}. In the same
vein, here we demonstrate that the same principle applies also to 
three-component (two IDP sequences plus solvent) systems. 
As described above, binary coexistence is governed by the free energy 
difference
\begin{equation}
\Delta f \equiv f_{\rm sep} - f_{\rm bulk} = v f(\phi_1^\alpha,\phi_2^\alpha) 
+ (1-v) f(\phi_1^\beta,\phi_2^\beta) - f(\phi_1^0,\phi_2^0) \; .
\end{equation}
Consider a modified free energy $g(\phi_1,\phi_2)$ with an additional
arbitrary linear function $a_0+a_1\phi_1+a_2\phi_2$ of $\phi$'s where
$a_0$, $a_1$, and $a_2$ are constants: 
\begin{equation}
g(\phi_1, \phi_2) = f(\phi_1,\phi_2) + a_0+a_1\phi_1+a_2\phi_2 \; .
\end{equation}
Now, the free energy difference between phase-separated and bulk phases becomes
\begin{equation}
\Delta g \equiv g_{\rm sep} - g_{\rm bulk} = 
\Delta f + a_1 \left[ v \phi_1^\alpha + (1\!-\!v) \phi_1^\beta - 
\phi_1^0 \right] +
a_2 \left[ v \phi_2^\alpha + (1\!-\!v)\phi_2^\beta -  \phi_2^0 \right]
	\label{eq:Delta_g}
\end{equation}
because $a_0$ in $g_{\rm sep}$ and $g_{\rm bulk}$ cancel. 
The two bracketed terms in Eq.~(\ref{eq:Delta_g}) are identically zero
because of Eq.~(\ref{eq:density_conserve}).
Hence $\Delta g = \Delta f$, meaning that any
linear function of $\phi$'s added to $f$, such as the
one utilized in Fig.~\ref{fig:f_2D3D} for graphical clarity, has no impact on
phase separation.

\begin{table}%[htbp]
   \centering
   \begin{tabular}{@{} cccccc @{}} % Column formatting, @{} suppresses leading/trailing space
      Sequence  & sv label & $-$SCD & 
$\kappa$
      & $R_{\rm g}$/\AA & $T_{\rm cr}^*$ \\
      \hline 
       seq1 &  sv28 & 15.99 & 0.7666 & 17.9 & 5.177 \\
       seq2 &  sv24 & 17.00  & 0.4456 & 17.6 & 5.160 \\
       seq3 &  sv25 & 12.77  & 0.5283 & 19.6 & 4.144 \\
       seq4 &    sv20 & 7.37  & 0.2721 & 19.6 & 2.275 \\
       seq5 &    sv15 & 4.35  & 0.1354 & 20.4 & 1.282 \\
       seq6 &    sv10 & 2.10 & 0.0834 & 25.5 & 0.611 \\
       seq7 &    sv1   & 0.41  & 0.0009 & 29.9 & 0.089 \\
      \hline \\
   \end{tabular}
   \caption{Sequences studied are identified as seq1--7 in
this article. They correspond to seven of the thirty 50-residue charged 
sequences with zero 
net charge in Das and Pappu \cite{Das13}. The sequences' sv labels, 
the values of their charge pattern parameter $\kappa$ and simulated 
single-chain 
radius of gyration $R_{\rm g}$ are 
those in the same reference \cite{Das13}.
The charge pattern parameter SCD is that of Sawle and Ghosh \cite{Sawle15}.
Values of SCD and the RPA-predicted critical  
temperature $T^*_{\rm cr}$ are from Lin and Chan \cite{Lin17b}.}
   \label{table:SCDofSeqs}
\end{table}

We apply the above-described minimization procedure for three-component
(two IDP sequences plus solvent) systems to determine $(\alpha, \beta)$ 
for selected pairs of sequences in Table~\ref{table:SCDofSeqs}. To minimize 
possible numerical errors, every
set of $\{\phi_1^\alpha,\phi_2^\alpha,\phi_1^\beta,\phi_2^\beta\}$
obtained by minimizing $f_{\rm sep}$ is subject to further testing
by comparing the chemical potentials in $\alpha$ and $\beta$. 
As described in Eq.~(A.5) of Ref.~\cite{Lin17a}, phase equilibrium 
implies the following equalities,
\begin{subequations}
\begin{align}
f_1^{'\alpha} & = f_1^{'\beta} \label{eq:chem_eq_df1} \\  
f_2^{'\alpha} & = f_2^{'\beta} \label{eq:chem_eq_df2}  \\ 
%f^\alpha - \phi_1^\alpha f_1^{'\alpha} - \phi_2^\alpha f_2^{'\alpha}  & = 
%f^\beta - \phi_1^\beta f_1^{'\beta} - \phi_2^\beta f_2^{'\beta}  
\mu_{\rm w}^\alpha & = \mu_{\rm w}^\beta
	\label{eq:chem_eq_f-df1-df2} 
\end{align} 
	\label{eq:chem_bal}%
\end{subequations}
where
\begin{equation}
\begin{aligned}
f^y \equiv & f(\phi_1^y, \phi_2^y) \; , \\
f_x^{'y} \equiv & \left. 
\frac{\partial f(\phi_1,\phi_2)}{\partial \phi_x}\right|_{(\phi_1,\phi_2)=(\phi_1^y, \phi_2^y)}, \\ 
x = & 1,2  \; ; \; y = \alpha,\beta \; ,
\end{aligned} 
\end{equation}
and 
\begin{equation}
\mu_{\rm w}^y \equiv f^y - \phi_1^y f_1^{'y} - \phi_2^y f_2^{'y}
\label{mu_def}
\end{equation}
is the chemical potential of water~\cite{Lin17a}. 
Making use of the volume conservation conditions in Eq.~(\ref{eq:chem_bal}) 
and substituting Eq.~(\ref{eq:phi_1_beta}) for $\phi_1^\beta$ and
Eq.~(\ref{eq:phi_2_beta}) for $\phi_2^\beta$, Eq.~(\ref{eq:chem_bal})
becomes three equalities for three variables $\phi_1^{\alpha}$,
$\phi_2^{\alpha}$, and $v$. It follows that a unique determination of the 
phase-separated volume fractions $\phi_1^\alpha$, $\phi_2^\alpha$, 
$\phi_1^\beta$, and $\phi_2^\beta$ is afforded by Eq.~(\ref{eq:chem_bal}).

It is straightforward to show that the set of phase-separated volume
fractions $\{\phi_1^\alpha,\phi_2^\alpha,\phi_1^\beta,\phi_2^\beta\}$ 
determined by Eq.~(\ref{eq:chem_bal}) are identical to that obtained
by minimizing $f_{\rm sep}$ in Eq.~(\ref{eq:f_sep_3var}). A necessary 
condition for the minimization of $f_{\rm sep}$
is that its Jacobian vector ${\bf J}_{\rm sep}$ of
first-order partial derivatives of independent variables vanishes:
\begin{equation}
{\bf J}_{\rm sep}(v, \phi_1^\alpha, \phi_2^\alpha) \equiv \left(
\footnotesize
\begin{aligned}
\frac{\partial f_{\rm sep}}{\partial \phi_1^\alpha} \\
\frac{\partial f_{\rm sep}}{\partial \phi_2^\alpha} \\
\frac{\partial f_{\rm sep}}{\partial v}  
\end{aligned}
\right) = {\bf 0} \; .
	\label{eq:Jsep=0}
\end{equation} 
In other words,
\begin{subequations}
\begin{align}
%\frac{\phi_1^o-\phi_1^\alpha}{1-v} f_1^{'\beta} + \frac{\phi_2^o-\phi_2^\alpha}{1-v} f_2^{'\beta}
\frac{\partial f_{\rm sep}}{\partial \phi_1^\alpha} = & v \left( f_1^{'\alpha} - f_1^{'\beta} \right) = 0  \; ,
	\label{eq:Jsep_dfsepdphi1a} \\
\frac{\partial f_{\rm sep}}{\partial \phi_2^\alpha} = & v \left( f_2^{'\alpha} - f_2^{'\beta} \right) = 0 \; ,
	\label{eq:Jsep_dfsepdphi2a} \\
\frac{\partial f_{\rm sep}}{\partial v}  = & f^\alpha - f^\beta + 
(\phi_1^\beta-\phi_1^\alpha) f_1^{'\beta} + (\phi_2^\beta-\phi_2^\alpha) f_2^{'\beta} = 0  \; ,
	\label{eq:Jsep_dfsepdv} 
\end{align}
	\label{eq:Jsep}%
\end{subequations} 
wherein we have utilized Eq.~(\ref{eq:phi_1_beta})
for $\phi_1^\beta$ and Eq.~(\ref{eq:phi_2_beta}) for $\phi_2^\beta$.
Clearly, Eqs.~(\ref{eq:Jsep_dfsepdphi1a}) and (\ref{eq:Jsep_dfsepdphi2a}) 
are equivalent to Eqs.~(\ref{eq:chem_eq_df1}) and (\ref{eq:chem_eq_df2}), 
respectively, and Eq~(\ref{eq:Jsep_dfsepdv}) is equivalent to 
Eq.~(\ref{eq:chem_eq_f-df1-df2}) by virtue of Eq.~(\ref{mu_def}). Q.E.D. 

Starting with $(\alpha, \beta)$ obtained by minimizing $f_{\rm sep}$
in Eq.~(\ref{eq:f_sep_3var}),
only those that deviate less than 0.1\% from the chemical-potential-balancing 
equalities in Eq.~(\ref{eq:chem_bal}) 
are accepted as valid binary pairs in our analysis. For the 
sequence pairs (seq1, seq5) and (seq1, seq6), a smaller threshold of
0.01\% is used to ensure accuracy of the computed phase-separated $\phi$'s
because for these sequence pairs the chemical potential balancing conditions 
are quite insensitive to variations of the $\phi$'s.

\subsection{Binary coexistence of two charged sequences}

Using RPA, we investigated previously how the phase separation behaviors 
of charged IDP sequences are affected by their charge 
patterns~\cite{Lin16,Lin17a,Lin17b}. In particular, for the set of thirty 
KE sequences of Das and Pappu~\cite{Das13} with zero net charge but an 
equal number of 25 positively charged lysine (K) and 25 negatively charged 
aspartic acids (E) in different permutations, the critical temperature
$T^*_{\rm cr}$ of phase separation was found~\cite{Lin17b} to be correlated 
with charge pattern parameters $\kappa$ \cite{Das13} and
\begin{equation}
\mathrm{SCD} \equiv \frac{1}{N}\sum_{i=1}^N\sum_{j=i+1}^N \sigma_i \sigma_j 
\sqrt{j-i} \; ,
\label{eq:SCD}
\end{equation}
where $i,j$ label the residues with charges $\sigma_i,\sigma_j$ along a chain 
of length $N$~\cite{Sawle15}. The $\kappa$ and SCD parameters exhibit 
similar correlations with single-chain radius of gyration 
$R_{\rm g}$~\cite{Das13, Sawle15}. 
The correlation of $T^*_{\rm cr}$ and $R_{\rm g}$ with SCD is stronger 
than that with $\kappa$. A likely reason is that
SCD accounts for nonlocal effects between charges far apart along the
chain sequence whereas $\kappa$ does not~\cite{Lin17b}.
Although we use only SCD in our analysis below, an equivalent analysis
using $\kappa$ is expected to produce a similar trend.

How does the phase behavior of an IDP solution with two sequences depend
on the sequences' difference in charge patterns? Intuitively, when 
two sequences with different SCD values are present together,
their different propensities to phase separate are expected
to interfere. Indeed, such an effect of inter-sequence
interference is seen clearly in the Taylor expansion of the integrand
of the RPA expression $f_{\rm el}$ for the electrostatic contribution 
to free energy in Eq.~(\ref{eq:fel_integral}), 
\begin{equation}
\begin{aligned}
\ln[1 + {\cal G}(k)] - {\cal G}(k) = & -\frac{1}{2}{\cal G}(k)^2 + \frac{1}{3}{\cal G}(k)^3 + ... \\
= & -\frac{1}{2}\langle \sigma_1 | \hat{G}_{11}^*(k) | \sigma_1 \rangle^2 
-\frac{1}{2}\langle \sigma_2 | \hat{G}_{22}^*(k) | \sigma_2 \rangle^2  \\
& - \langle \sigma_1 | \hat{G}_{11}^*(k) | \sigma_1 \rangle
	\langle \sigma_2 | \hat{G}_{22}^*(k) | \sigma_2 \rangle \\
& + O\left({\cal G}(k)^3\right) \; ,
\end{aligned}
	\label{eq:G_expand}
\end{equation}
where $\hat{G}_{11}^*(k)$ and $\hat{G}_{22}^*(k)$
are the product of $4\pi/[k^2(1+k^2)T^*]$ with, respectively, the
$\hat{G}_{11}(k)$ and $\hat{G}_{22}(k)$ in Eq.~(\ref{calG_def}).
The first two terms after the second equality in Eq.~(\ref{eq:G_expand}) are
self-interactions of the two sequences, identical to those in one-sequence 
RPA theory (see, e.g. Eq.~(1) in Ref.~\cite{Lin17b}). The third 
term represents the interference effect in RPA. Since it is
the product of square roots of the two self-interaction terms, 
its strength is intermediate between them, suggesting that
phase behaviors of two-sequence systems are sensitive to the
similarity/dissimilarity in charge pattern between the two sequences.

\begin{figure}%[htbp] 
\includegraphics[width=\columnwidth]{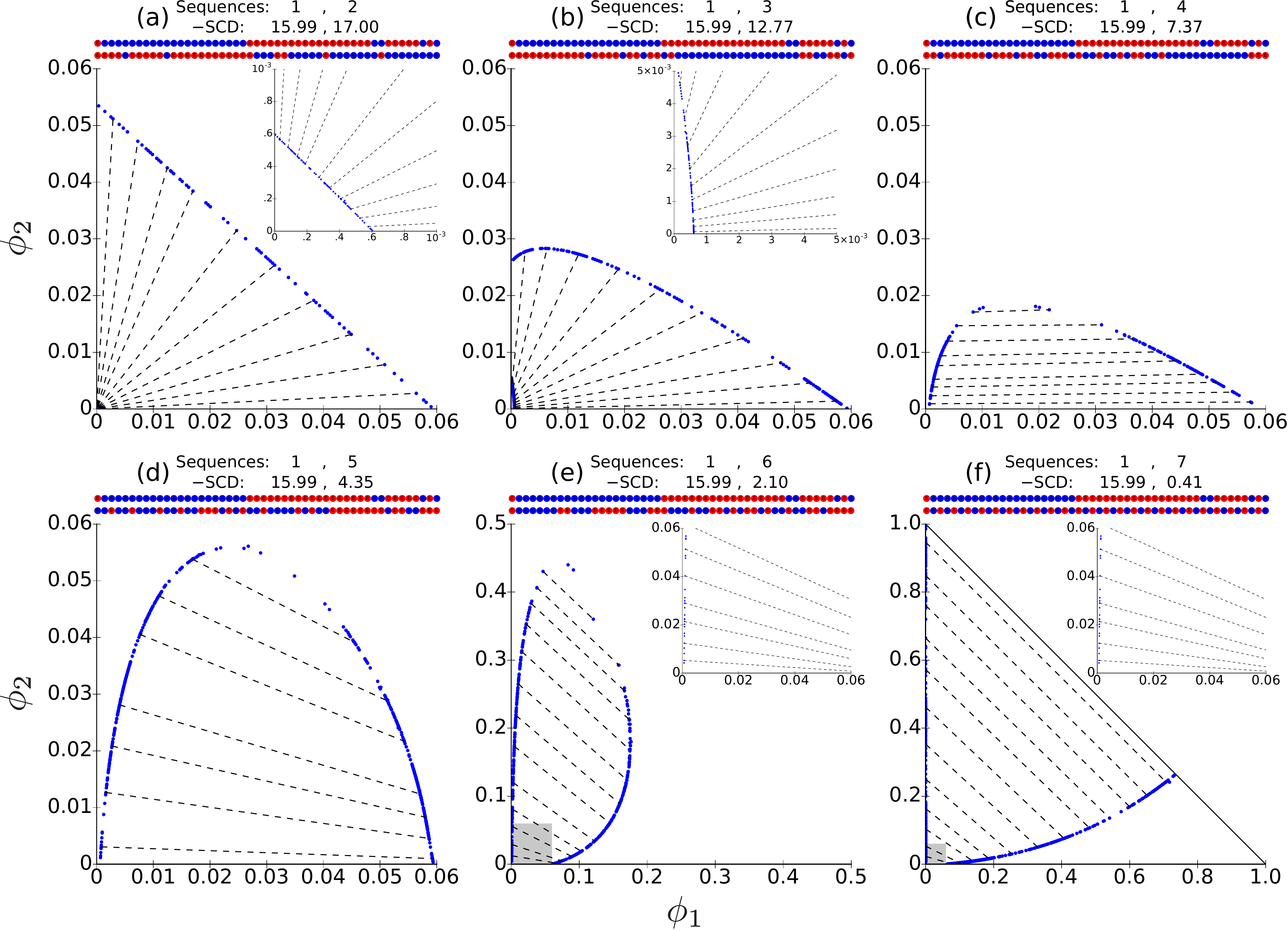} 
\caption{RPA-predicted binary phase diagrams of six charged sequence pairs at 
$T^* = 4$. In each panel, the pair of sequences considered and their
charge pattern parameters ($-$SCD) are shown at the top, with
positively and negatively charged residues depicted as red
and blue circles respectively. Further properties of the sequences are 
provided in Table~\ref{table:SCDofSeqs}. Horizontal axes
($\phi_1$) refer to the volume fraction of seq1, vertical axes ($\phi_2$)
are for the volume fraction of the other sequence in each of the pairs. 
The charge patterns of the 
sequence pairs vary from being very similar (a) to very dissimilar (f).
This trend is quantified by the difference in $-$SCD values between the 
two sequences in each of the six pairs.
Each dashed line is a tie line connecting a pair of blue dots that 
represent coexisting phases $\alpha=(\phi_1^\alpha,\phi_2^\alpha)$ and 
$\beta=(\phi_1^\beta,\phi_2^\beta)$.
All bulk volume fractions $(\phi_1^0,\phi_2^0)$ lying on one tie line 
undergo phase separation to the same $(\alpha, \beta)$; and, following
Eq.~(\ref{eq:density_conserve}), $v/(1-v)=$
$(\phi_1^\beta-\phi_1^0)/(\phi_1^0-\phi_1^\alpha)=$
$(\phi_2^\beta-\phi_2^0)/(\phi_2^0-\phi_2^\alpha)$, which is
equal to the length ratio of the two segments of the tie line 
from the bulk volume fractions $(\phi_1^0,\phi_2^0)$ 
on the phase diagram to the phase boundary marked by the blue dots.
The black inclined solid line in (f), $\phi_1+\phi_2=1$, delimits the 
region $\phi_1+\phi_2\le 1$ within which IDP volume fractions may vary.
Insets in (a) and (b) are zoom-in plots of a part of the phase
diagram with extremely low $\phi_1$ and $\phi_2$. They offer
a clearer view of the dilute phase boundaries for
the (seq1, seq2) and (seq1, seq3) pairs.
Insets in (e) and (f) are zoom-in plots of the 
grey-shaded regions of the respective phase diagrams. 
To provide a scale for comparison, the size of the grey-shaded regions
$\phi_1,\phi_2 \in [0,0.06]$ in (e) and (f) is chosen to be equal
to the plotted regions for the other four sequence pairs.
}
   \label{fig:6biCoex}
\end{figure}

To investigate this sensitivity, we use the sequences in 
Table~\ref{table:SCDofSeqs} to compute the phase diagrams of
six pairs of sequences, namely seq1 with each of the six other
sequences. The pairs are selected to represent a broad range of 
similarity/dissimilarity in charge pattern as quantified by
the difference in SCD values: from the (seq1, seq2) pair with 
SCD $=(-15.99,-17.00)$ to (seq1, seq7) with SCD $=(-15.99,-0.41)$. 
To compare the phase behaviors of the six sequence pairs on an
equal footing, all phase diagrams in Fig.~\ref{fig:6biCoex}
are computed at the same reduced temperature $T^*=4$. Noting that
condensed-phase volume fractions tend to decrease with increasing
$T^*$, this temperature is chosen because it falls in the mid-range 
of the broad span of $T^*_{\rm cr}$'s for the sequences in
Table~\ref{table:SCDofSeqs}. $T^*=4$ is much higher than 
the $T^* = 0.55$ equivalent of room temperature ($T = 300$ K) when an aqueous
$\epsilon_{\rm r}=80$ is assumed~\cite{Lin17b}. This seemingly unphysical 
condition in our calculation has little impact, however, on the present 
goal of ascertaining general principles and behavioral trends.
Although we do not aim for direct, detailed comparison with experiment here,
$T^*=4$ is experimentally relevant, for
example, to IDPs with charge patterns similar to those considered here but 
with their electrostatic interaction strength significantly scaled down for
various physical reasons such as screening or a more sparse charge
distribution along the IDP sequence.

%%%%%%%%%%%%%%%%%%%%%%%%%%%%%%%%%%%%%%%%%%%%%%%%%%%%%%%%%%%%%%%%%%%%%%%%%%%%%%

For each of the six phase diagrams in Fig.~\ref{fig:6biCoex}, the area 
surrounded by blue dots and ``shaded'' by black dashed lines
is the region of binary coexistence. In other words,
bulk-state volume fractions falling within this region will phase
separate into two coexisting liquid phases [as in Fig.~1(c)--(e)],
whereas bulk-state volume 
fractions residing outside this region will be stable as a single 
liquid phase [as in Fig.~1(b)]. 
Every black dashed line is the tie line connecting a 
pair of blue dots representing separated phases $(\alpha, \beta)$ for 
any bulk-state volume fractions lying on the given tie line
[corresponding to the arrows in the schematic
phase digrams for Fig.~1(c)--(e)]. 

The average slope of the tie lines changes from positive for similarly
patterned sequences [Fig.~\ref{fig:6biCoex}(a), (b)] to negative for 
very differently patterned sequences [Fig.~\ref{fig:6biCoex}(e), (f)].
This trend may be understood as follows.
When both sequences of a given pair can undergo phase separation
individually ($T^*_{\rm cr}>4$ for both), the tie lines near the
$\phi_1$ and $\phi_2$ axes must be close to being parallel
to the axes because when either $\phi_1^0\to 0$ or $\phi_2^0\to 0$, 
the two-sequence system reduces to the corresponding single-sequence system
that phase separates. This situation applies to (seq1, seq2) and 
(seq1, seq3), resulting in positive tie-line slopes, indicating that 
the populations of
the two sequences in each pair are well mixed even when they undergo
phase separation. They prefer to stay together after they phase separate, 
with similar population ratios for the two sequences 
in the ``both-dilute'' (small $\phi$'s) as well as the ``both-condensed''
(larger $\phi$'s) phases [as in Fig.~1(c)].

Because $T^*_{\rm cr} < 4$ for the other four sequences (seq4--seq7), 
they do not phase separate by themselves individually 
and therefore tie lines near
the $\phi_2$-axis need not be approximately parallel to it. Nonetheless,
tie lines close to the $\phi_1$-axis are still required to essentially
line up with the axis. For tie lines that possess large $\phi_2$ values, the 
volume conservation condition $\phi_1+\phi_2=1$ enforces negative tie-line
slopes. The combined effect of these constraints lead to tie-line
slopes that gradually change from $\approx 0$ near the $\phi_1$-axis 
to $\approx -1$ near the $\phi_1+\phi_2=1$ boundary, as exemplified
by the case of (seq1, seq7) in Fig.~\ref{fig:6biCoex}(f). As shown
in Fig.~\ref{fig:6biCoex}(d) and (e) for (seq1, seq5) and 
(seq1, seq6), this trend is apparent even when the phase-separated 
regime does not extend all the way to the $\phi_1+\phi_2=1$ boundary.
Negative tie-line slopes imply various degrees of demixing of the
populations of the two
sequences: the phase-separated state $(\alpha,\beta)$ now comprises one 
$\phi_1$-enriched ($\phi_1^\alpha \gg \phi_2^\alpha$) phase coexisting
with one $\phi_2$-enriched ($\phi_2^\beta \gg \phi_1^\beta$) phase.
The degree of population demixing depends on how dissimilar are the
charge patterns of the two sequences in the pair. 
For large differences in SCD as in Fig.~\ref{fig:6biCoex}(f),
one of the coexisting phases can have a very low population of seq1 
but a substantial seq7 volume fraction, whereas the other
phase has a relatively low population of seq7 but a substantial
seq1 volume fraction [as in Fig.~1(e)].

The (seq1, seq4) pair in Fig.~\ref{fig:6biCoex}(c) is at
the crossover between the well-mixed and demixed extremes.
Tie-line slopes in this case are all $\approx 0$, indicating
that although increasing seq4 volume fraction decreases the phase 
separation tendency of seq1, even in the phase-separated regime the 
concentration of seq4 is essentially identical in the two coexisting phases,
i.e., $\phi_2^\alpha \simeq \phi_2^\beta \simeq \phi_2^0$ [as in Fig.~1(d)].

%%%%%%%%%%%%%%%%%%%%%%%

\begin{figure}%[htbp] 
\centering
\includegraphics[width=0.5\columnwidth]{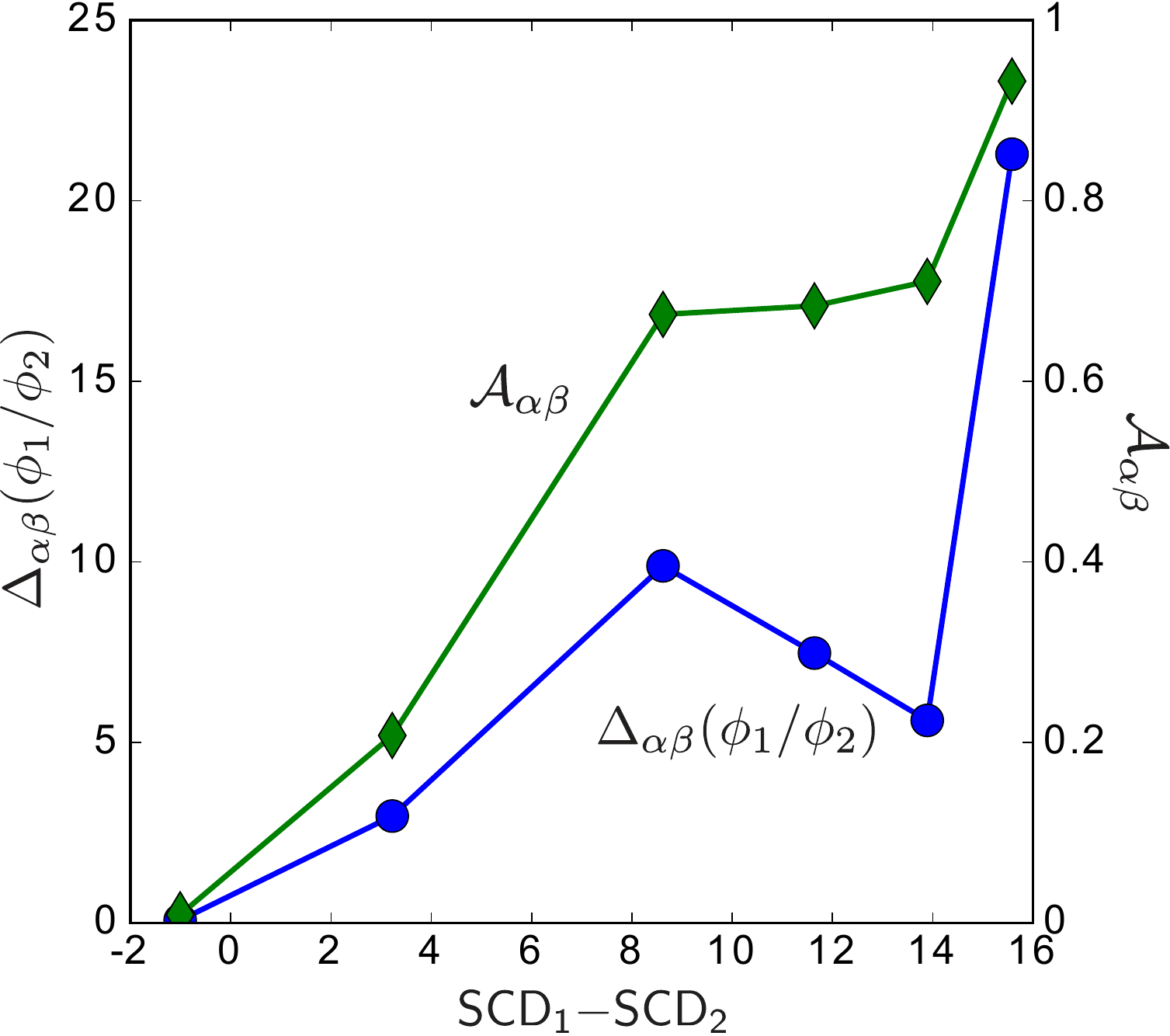} 
\caption{Trend of the component ratio $\phi_1/\phi_2$ in coexisting phases $(\alpha,\beta)$ for different pairs of charged sequences. Two measures, 
$\Delta_{\alpha\beta}\left( \phi_1/\phi_2 \right)$ and 
${\cal A}_{\alpha\beta}$, are plotted. When the charge patterns of the
sequences in a pair are similar (small SCD$_1-$SCD$_2$), 
the two sequences tend to be well-mixed with similar relative volume
fractions in a ``both-dilute" phase and a ``both-condensed" phase. 
When the charge patterns of the sequences in a pair are dissimilar 
(large SCD$_1-$SCD$_2$), the two sequences tend to demix in that they 
largely exclude each other in the two separated phases. 
This mixing/demixing behavior is less extreme for intermediate 
SCD$_1-$SCD$_2$ values.
}
   \label{fig:SCD_phi_ratios}
\end{figure}

%%%%%%%%%%%%%%%%%%%%

The difference in the ratio of component populations in coexisting
phases $(\alpha,\beta)$ may be quantified by comparing the 
volume ratio of the two sequences in the two phases. We first
consider a rather intuitive measure 
\begin{equation}
\Delta_{\alpha\beta}\left( \phi_1/\phi_2 \right) \equiv \left\langle 
\left| \frac{\phi_1^\alpha}{\phi_2^\alpha} - \frac{\phi_1^\beta}{\phi_2^\beta} \right| 
\right\rangle \; 
\end{equation}
of compositional asymmetry between coexisting phases,
where the absolute value ensures that $\Delta_{\alpha\beta}(\phi_1/\phi_2)$
is $\alpha\leftrightarrow\beta$ symmetric, and the bracket 
$\langle\dots\rangle$ denotes averaging over all 
$(\alpha, \beta)$ pairs of coexisting phases. 
One disadvantage of this measure, however, is that the average is strongly
dominated by those pairs of $(\alpha, \beta)$ with large $\phi_1$ but 
small $\phi_2$, i.e., coexisting pairs that are close to $\phi_1$-axis 
in Fig.~\ref{fig:6biCoex}.
Therefore, we also consider another composition asymmetry measure 
\begin{equation} 
{\cal A}_{\alpha\beta} \equiv \left\langle \frac{2}{\pi}\left| 
\tan^{-1}\left(\frac{\phi_1^\alpha}{\phi_2^\alpha}\right) - 
\tan^{-1}\left(\frac{\phi_1^\beta}{\phi_2^\beta}\right) \right| \right\rangle
\;
\end{equation}
that avoids this potentially problematic feature by replacing
the ratio $\phi_1/\phi_2$ with its arctangent value normalized by $\pi/2$
such that $0\le {\cal A}_{\alpha\beta}\le 1$. 

Summarizing our findings using two-sequence RPA theory,
Fig.~\ref{fig:SCD_phi_ratios} shows the
variation of $\Delta_{\alpha\beta}(\phi_1/\phi_2)$ as well as
${\cal A}_{\alpha\beta}$ with the difference in SCD values
of the six sequence pairs in Fig.~\ref{fig:6biCoex}. 
A reasonable correlation is seen for both composition asymmetry measures, 
with the ${\cal A}_{\alpha\beta}$ measure exhibiting a better correlation 
by varying monotonically with SCD difference, indicating that compositional 
asymmetry or degree of demixing of phase-separated populations 
as quantified by ${\cal A}_{\alpha\beta}$ 
is positively correlated with the difference in charge patterns as 
quantified by difference in SCD values.
This plot illustrates graphically how a stochastic, multivalent form
of molecular recognition that arises from the interactions among
the diverse conformations in a multiple-chain ensemble can lead to 
demixing of different IDP species into different coexisting phases.

\section{Comparison with phase separation in Flory-Huggins (FH) models}

We next seek a deeper understanding of the RPA results and
their ramifications by comparing them with the predictions of 
a variety of FH models.
As emphasized, unlike RPA, FH by itself does not address the physics 
of sequence dependence \cite{Nott15,Lin17a}. 
Accordingly, FH $\chi$ interaction parameters
for IDP sequences have to be provided phenomenologically by experiment 
or theoretically by microscopic physical 
theory.\footnote{In the caption describing the FH results in Fig.~8
of Ref.~\cite{Lin17a}, $r_d$ is in fact the symbol $r$ for
the equilibrium spacing in Eq.(S19) of Ref.~\cite{Nott15}. This
typographical error does not affect the results. 
It should also be noted that because Ref.~\cite{Nott15} equates 
the ionic strength $I$
with [NaCl] but not 2[NaCl], their effective Debye length is 4.3~\AA~
instead of the correct value of 3.04~\AA. To facilitate comparison 
with Ref.~\cite{Nott15}, however,
the effective Debye length in Fig.~8 of Ref.~\cite{Lin17a} was also 
set to 4.3~\AA.} 
For example, as will be 
discussed further below, an intuitive and semi-quantitative connection 
between RPA and FH is provided by the expansion in Eq.~(\ref{eq:G_expand}).
It should also be noted that FH neglects interaction terms that are
higher than quadratic order in IDP volume fractions/concentrations 
($\phi$'s) such as the $O({\cal G}(k)^3)$ terms in Eq.~(\ref{eq:G_expand})
because ${\cal G}(k)\propto \phi$ [Eqs.~(\ref{G_eqsa}) and (\ref{G_eqsb})]. 
This approximation can be problematic when IDP concentrations are high.
Nonetheless, by treating the three parameters $\chi_{11}$, $\chi_{22}$, 
and $\chi_{12}$ in the three-component FH interaction term for
two IDP species plus solvent
\begin{equation}
f_{\rm int}^{\rm FH} = -\left( \chi_{11} \phi_1^2 + \chi_{22} \phi_2^2 + 
2\chi_{12}\phi_1\phi_2 \right) \; 
	\label{eq:fint_FH_3chi}
\end{equation}
as free (arbitrary) variables, we can either match FH behavior to
that of RPA to gain conceptual insights or explore other interaction 
scenarios that might be physically plausible when interactions 
other than the rudimentary electrostatics embodied in RPA are included 
in the physical picture.

\subsection{FH models that imitate RPA theory by having two independent 
$\chi$'s}\label{sec:imitate}

\begin{figure}%[htbp] 
\includegraphics[width=\columnwidth]{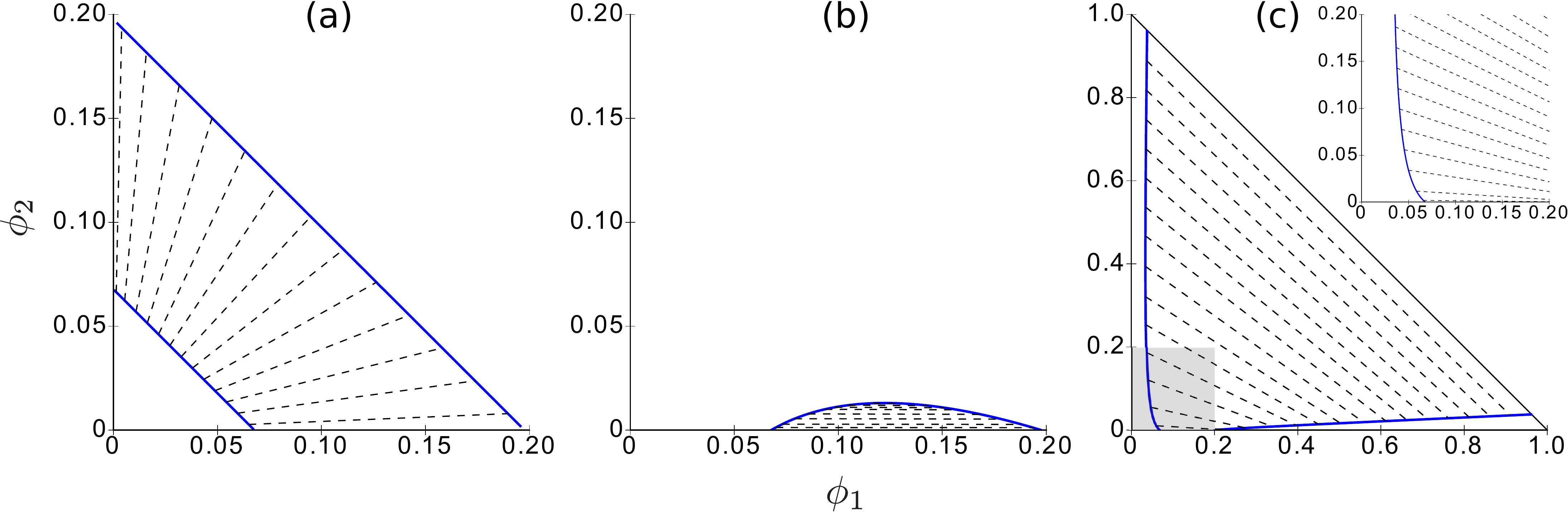} 
\caption{Binary phase diagrams of three-component FH systems of 
$N_1=N_2=50$ satisfying the RPA-like condition 
$\chi_{12} = \sqrt{\chi_{11}} \sqrt{\chi_{22}}$: 
(a) $\chi_{11} = \chi_{22} = 0.66$, 
(b) $\chi_{11} = 0.66$, $\chi_{22} = 0.5$, and
(c) $\chi_{11} = 0.66$, $\chi_{22} = 0.3$.
The shaded area in (c) indicates the region covered by
the inset as well as the entire plotted regions of (a) and (b).
As in Fig.~\ref{fig:6biCoex}, each dashed line is a tie line connecting 
a pair of coexisting phases depicted in blue.
The inclined black solid line in (c) is
the same $\phi_1+\phi_2=1$ volume-conservation boundary as that
shown in Fig.~\ref{fig:6biCoex}(f).
}
   \label{fig:3FH_line}
\end{figure}

We begin this analysis by first constructing FH models with interaction 
schemes similar to RPA, then tuning the interaction parameters to produce 
phase behaviors similar to those predicted by RPA in Fig.~\ref{fig:6biCoex}.
If we identify the $\int dk\;k^2/4\pi^2$ integral [Eq.~(\ref{eq:fel_integral})] 
of the order ${\cal G}(k)^2$ terms in the RPA expansion
Eq.~(\ref{eq:G_expand}) with the FH interaction term in 
Eq.~(\ref{eq:fint_FH_3chi}), 
we may define
$\tilde{\chi}_{11}(k)\phi_1^2\equiv
2\{\langle \sigma_1 | \hat{G}_{11}(k)|\sigma_1\rangle/[ k (1+k^2) T^*]\}^2$
and
$\tilde{\chi}_{22}(k)\phi_2^2\equiv
2\{\langle \sigma_2 | \hat{G}_{22}(k)|\sigma_2\rangle/[ k (1+k^2)T^*]\}^2$
such that
$\chi_{11}=\int dk\; \tilde{\chi}_{11}(k)$,
$\chi_{22}=\int dk\; \tilde{\chi}_{22}(k)$, and
$\chi_{12}=
\int dk\; \sqrt{\tilde{\chi}_{11}(k)}\sqrt{\tilde{\chi}_{22}(k)}$,
from which it is clear that only two set of interaction parameters
$\tilde{\chi}_{11}(k)$ and $\tilde{\chi}_{22}(k)$ as functions of $k$
are independent. Here we approximate this dependence by 
constraining $\chi_{12} = \sqrt{\chi_{11}} \sqrt{\chi_{22}}$.

We then construct three FH systems that have
$\chi_{11} = \chi_{22}$, $\chi_{11} \gtrsim \chi_{22}$, and 
$\chi_{11} \gg \chi_{22}$, corresponding respectively to 
sequence pairs with small, intermediate, and large charge pattern 
(SCD) differences. The phase diagrams of
these models (Fig.~\ref{fig:3FH_line}) exhibit a trend similar
to that seen in the RPA-predicted Fig.~\ref{fig:6biCoex}.
Specifically, Fig.~\ref{fig:3FH_line}(a) is similar to 
Fig.~\ref{fig:6biCoex}(a), Fig.~\ref{fig:3FH_line}(b) 
to Fig.~\ref{fig:6biCoex}(c), and Fig.~\ref{fig:3FH_line}(c) 
to Fig.~\ref{fig:6biCoex}(f). This correspondence offers conceptual
clarity because the degree to which the interaction between 
the two IDP species is favorable is explicit in FH.
When $\chi_{11} = \chi_{22}=\chi_{12}$, the two species are miscible and
their phase separation propensities are identical, resulting in
the coexistence of one both-dilute phase and one both-condensed phase.
The similarity between Fig.~\ref{fig:3FH_line}(a) 
Fig.~\ref{fig:6biCoex}(a) indicates that this behavior  
can be achieved physically by two IDP species with similar charge patterns.
In contrast, when $\chi_{11} \gg \chi_{12} \gg \chi_{22}$, miscibility
of the two species is poor and their phase separation propensities are 
quite different, resulting in a high degree of population demixing.  
The similarity
of this behavior shown in Fig.~\ref{fig:3FH_line}(c) with that in
Fig.~\ref{fig:6biCoex}(f) underscores once again that charge-pattern
mismatches between IDPs can lead to substantially weakening of attractive 
interactions.

\subsection{FH models with three independent $\chi$'s}

We next consider the general case in which the
three $\chi$'s in Eq.~(\ref{eq:fint_FH_3chi}) are independent. Although
this modeling setup does not have a simple correspondence with 
IDP sequences interacting via physical forces like that described 
above, the expanded variety of scenarios explored here would be valuable 
when behaviors much more complex than those allowed by our
current RPA formulation are considered
in more comprehensive and detailed physical theories.
As simple examples of the rich possibilities,
here we focus on FH models with $\chi_{11}=\chi_{22}$
($\equiv\chi$) and variable $\chi_{12}$ values that are not related to $\chi$. 

\begin{figure}%[htbp] 
\includegraphics[width=\columnwidth]{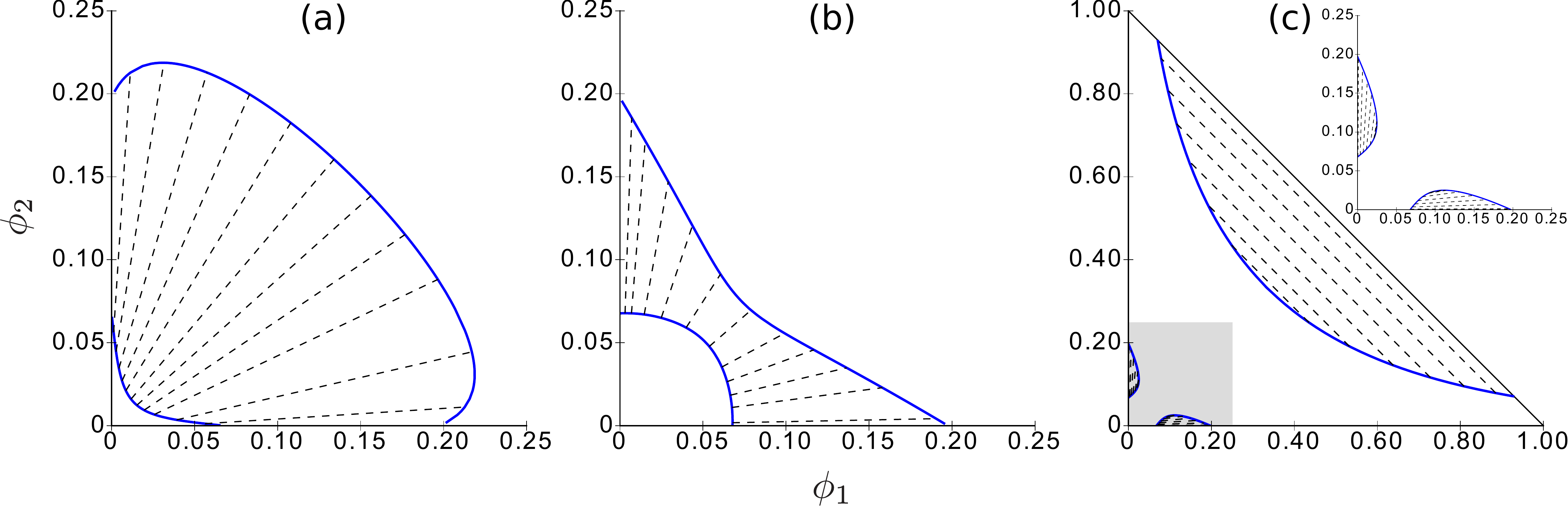} 
\caption{Binary phase diagrams of three-component FH systems of $N_1=N_2=50$,
$\chi\equiv\chi_{11}=\chi_{22}=0.66$, but with different $\chi_{12}$:
(a) $\chi_{12}= 0.72$,
(b) $\chi_{12} = 0.645$, and
(c) $\chi_{12}=0.63$.
The phase diagrams are $\phi_1\leftrightarrow \phi_2$ symmetric because
$\chi_{11}=\chi_{22}$.
The shaded area in (c) indicates the region covered by the inset as well 
as the entire plotted regions of (a) and (b).
Dashed lines and blue curves in (a), (b), and (c) and black inclined 
solid line  in (c) carry the same meanings as in Fig.~\ref{fig:6biCoex}(f).
}
   \label{fig:3FH_arbitrary}
\end{figure}

Fig.~\ref{fig:3FH_arbitrary} shows the phase diagrams of three representative 
models with (a) $\chi_{12} \gtrsim \chi$, (b) $\chi_{12} \lesssim \chi$, and 
(c) $\chi_{12} < \chi$. 
Compared to the $\chi\equiv\chi_{11}=\chi_{22}=\chi_{12}$ 
case in Fig.~\ref{fig:6biCoex}(a), it is clear that
a stronger inter-component attraction (larger $\chi_{12}$)
makes the phase-separated region bulge [Fig.~\ref{fig:3FH_arbitrary}(a)],
whereas a weaker inter-component attraction (smaller $\chi_{12}$)
shrinks it [Fig.~\ref{fig:3FH_arbitrary}(b)]. 
Nonetheless, the tie-line slopes are positive in both situations, indicating
that the two components are largely miscible.

Fig.~\ref{fig:3FH_arbitrary}(b) indicates that a weakened $\chi_{12}$ 
shrinks the phase-separated region most around $\phi_1=\phi_2$.
As $\chi_{12}$ decreases further, inter-component attraction
all but vanishes, micibility disappears, resulting
in the phase-separated region being broken into two parts, 
one for $\phi_1 \gg \phi_2$ and the other for $\phi_1 \ll \phi_2$
[Fig.~\ref{fig:3FH_arbitrary}(c), shaded area and inset].
The tie lines in these two regions are almost parallel to either
the $\phi_1$- or the $\phi_2$-axis, implying that one
component is dominant while the concentration of the other component
barely changes upon phase separation. 
With such an {\it effective} inter-component repulsion
(i.e., less favorable inter-component attraction
vis-\`a-vis the strengths of intra-component cohesion), an 
additional phase-separated regime of poor miscibility similar
to that in Figs.~\ref{fig:6biCoex}(f) and~\ref{fig:3FH_line}(c)
is induced, wherein all tie-line slopes are negative, 
signaling substantial demixing
[Fig.~\ref{fig:3FH_arbitrary}(c), region close to $\phi_1+\phi_2=1$ 
with tie-line slopes $=-1$]. 

\subsection{Ternary coexistence in FH model}\label{sec:ternaryFH}

If $\chi_{12}$ is made even weaker than that in 
Fig.~\ref{fig:3FH_arbitrary}(c), the region of poor miscibility 
below the $\phi_1+\phi_2=1$ boundary would grow. Finally, the
three phase-separated regions intersect and a new ternary 
coexistence region emerges in-between.

According to the Gibbs phase rule, an $n$-component system can separate into at 
most $n$ coexisting 
phases, when all other environmental conditions, e.g. temperature and
pressure, are kept constant~\cite{Lin17a,Atkins09}. In our system of $n=3$ (two
sequences plus solvent), whether the system will separate to two or three
coexisting
phases may be deduced by observing the variation of tie-line slopes in 
putative regions of binary coexistence: If the tie lines have to become
parallel to the $\phi_1$-axis, $\phi_2$-axis, or the
$\phi_1+\phi_2=1$ line when they approach these boundaries respectively,
the tie-line slopes have to be able to vary smoothly to satisfy these 
constraints in order for binary coexistence to be stable. In that case,
ternary coexistence is unlikely. Conversely, if there are conflicts
that prevent a smooth change of tie-line slope, a ternary coexistence 
region ensues.

\begin{figure}%[htbp] 
\centering
\includegraphics[width=\columnwidth]{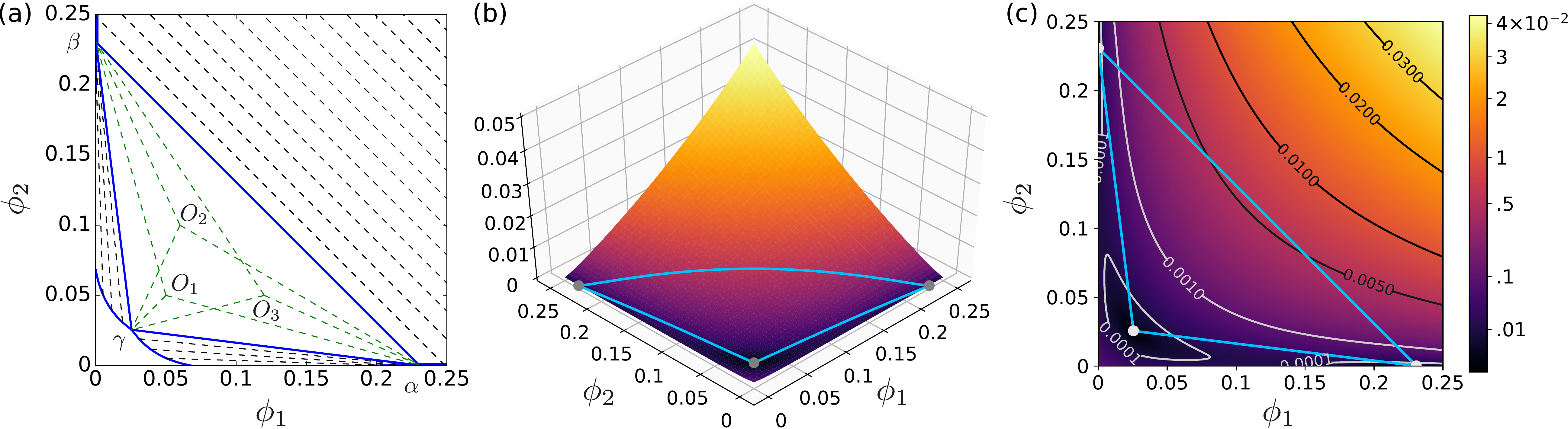} 
\caption{Ternary phase diagram of a three-component (two IDP species
plus solvent) FH system with IDP sequence lengths $N_1=N_2=50$, 
$\chi_{11}=\chi_{22}=0.66$, and $\chi_{12}=0.33$. (a) All bulk states
in the triangular region, labeled as $O$'s, undergo ternary phase 
separation to the three coexisting phases $\alpha$, $\beta$, and $\gamma$
at the vertices of the triangle [as in Fig.~1(f)].
Outside of the triangle there are three binary coexistence regions that 
converge to the $\phi_1$-axis, $\phi_2$-axis, and the $\phi_1+\phi_2=1$ 
boundary, respectively, wherein dashed tie lines connect
pairs of coexisting phases as in Figs.~\ref{fig:6biCoex}, \ref{fig:3FH_line}, 
and \ref{fig:3FH_arbitrary}. Bulk states within the bottom-left small white 
region near the $\phi_1$-$\phi_2$ origin does not phase separate.
The green dashed lines connect three examples of bulk state $O_1$, 
$O_2$, $O_3$ to the same ternary phases $\alpha$, $\beta$, $\gamma$ 
because the compositions of these phases are independent of bulk-state
volume fractions. For any given bulk state $O$, $v_\alpha$, $v_\beta$,
and $v_\gamma$ are given, respectively, by the areas of the triangles 
$O\beta\gamma$, $O\alpha\gamma$, and $O\alpha\beta$ (all have one solid 
blue and two green dashed sides) as a fraction of the total triangular area of 
$\alpha\beta\gamma$ bound by three solid blue lines (see text for details).
(b) Free energy landscape of $\Delta f(\phi_1,\phi_2) \equiv
f_{\rm bulk}(\phi_1,\phi_2) - [a_0+a_1\phi_1+a_2\phi_2]$,
where the bracketed function linear in the $\phi$'s specifies the plane
defined by $\alpha$, $\beta$, and $\gamma$. In other words, the $a$
coefficients are determined by solving $f_{\rm bulk}(\phi_1^z,\phi_2^z)
= a_0+a_1\phi_1^z+a_2\phi_2^z$ for $z=\alpha,\beta,\gamma$.  Note that 
$\Delta f > 0$ for all $(\phi_1,\phi_2)$ in the plotted region, indicating
that phase separation is preferred if the volume conservation condition for
either binary [Eq.~(\ref{eq:density_conserve})] or ternary 
[Eq.~(\ref{eq:3density})] coexistence can 
be satisfied.  The logarithmic color scale on the right is for both (b) 
and (c). The turquoise lines in (b) and (c) mark the same ternary phase 
boundaries as those in (a). 
(c) Contour plot of $\Delta f$. The three ternary coexisting phases
($\Delta f = 0$) are seen to be situated in three
different basins with small $\Delta f$.  }
   \label{fig:3FH_ternary}
\end{figure}

An example is provided by using Fig.~\ref{fig:3FH_arbitrary}(c) as
starting point. Here the three phase-separated regions are close to
the three boundaries, and their tie lines are essentially parallel to
the respective boundaries. Under this circumstance, when effective
inter-component repulsion is enhanced by weakening $\chi_{12}$ to
cause the three regions to evolve toward merging, the conflict among the
three different trends of tie-line slopes necessitates reconcilation by
a region of ternary phase separation (Fig.~\ref{fig:3FH_ternary}). 

In order to determine the three phases in ternary coexistence mathematically, 
we extend the phase-separated free energy expression
in Eq.~(\ref{eq:f_2sep}) for phases $(\alpha,\beta)$ to including one 
additional phase $\gamma$, viz.,
\begin{equation}
f_{\rm ternary} = 
v_\alpha f(\phi_1^\alpha,\phi_2^\alpha) 
+ v_\beta f(\phi_1^\beta,\phi_2^\beta)
+ (1\!-\!v_\alpha\!-\!v_\beta) f(\phi_1^\gamma,\phi_2^\gamma) \; ,
	\label{eq:f_ternary}
\end{equation}
where the fractional volumes $v_\alpha, v_\beta$ are functions of 
the six $\phi$'s by virtue of volume conservation,
\begin{equation}
\sum_y v_y \phi_x^y + \left(1- \sum_y v_y\right) \phi_x^\gamma=\phi_x^0,
\label{eq:3density}
\end{equation}
where $x=1,2$ and $\sum_y$ is over $y=\alpha,\beta$. 
Now the equalities in Eq.~(\ref{eq:chem_bal}) have to include the 
addition phase to become
\begin{subequations}
\begin{align}
f_1^{'\alpha} & = f_1^{'\beta} = f_1^{'\gamma} \; ,  \\  
f_2^{'\alpha} & = f_2^{'\beta} = f_2^{'\gamma} \; ,   \\ 
\mu_{\rm w}^\alpha & = \mu_{\rm w}^\beta = \mu_{\rm w}^\gamma \; .
\end{align} 
	\label{eq:ternary_equality}%
\end{subequations}
Because here we have six equalities in Eq.~(\ref{eq:ternary_equality}) 
for six phase-separated $\phi_x^y$'s, the solution for ternary
coexistence of $(\alpha, \beta, \gamma)$ is unique irrespective of
the bulk-state volume fractions insofar as they fall within the ternary
coexistence region. This situation is different from that of binary 
coexistence in which the separated phases $(\alpha, \beta)$ can be different
for different bulk-state $(\phi_1^0,\phi_2^0)$'s when they are
on different tie lines.

Similar to the binary coexistence case in Sec.~\ref{sec:Binary_3comp},
we proceed to demonstrate that minimizing Eq.~(\ref{eq:f_ternary}) 
is equivalent to solving the equations in Eq.~(\ref{eq:ternary_equality}). 
Using essentially the same approach, we rewrite $f_{\rm ternary}$ as 
a function of six independent variables: $v_\alpha$, $v_\beta$, 
$\phi_1^\alpha$, $\phi_2^\alpha$, $\phi_1^\beta$, and $\phi_2^\beta$
by first utilizing Eq.~(\ref{eq:3density}) to express $\phi_1^\gamma$ 
and $\phi_2^\gamma$ as
\begin{equation}
\phi_x^\gamma = \frac{\phi_x^0-\sumt_y v_y \phi_x^y}{1- \sumt_y v_y} \; ,
\end{equation}
where $x,y$ and $\sum_y$ have the same meanings as above. In this
notation, Eq.~(\ref{eq:f_ternary}) becomes
\begin{equation}
f_{\rm ternary} 
(v_\alpha,v_\beta,\phi_1^\alpha,\phi_2^\alpha,\phi_1^\beta,\phi_2^\beta)
= \sum_y v_y f^y + \left(1-\sum_y v_y\right) f^\gamma \; .
\end{equation}
Similarly to Eq.~(\ref{eq:Jsep=0}) for binary phase separation, we 
calculate the six derivatives of $f_{\rm ternary}$ and set them
to zero as in Eq.~(\ref{eq:Jsep}) as necessary conditions for
the minimization of $f_{\rm ternary}$, resulting in
\begin{subequations}
\begin{align}
\frac{\partial f_{\rm ternary} }{\partial \phi_x^y} = & v_y f^{' y}_x 
- \left(1-\sum_{y^\prime} v_{y^\prime} \right) 
f_x^{'\gamma} \cdot \frac{-v_y}{ 1-\sumt_{y^\prime} v_{y^\prime}} 
= v_y \left( f^{' y}_x - f_x^{'\gamma} \right) = 0  \; ,
\label{eq:dfter_dphiyx} \\
\frac{\partial f_{\rm ternary} }{\partial v_y} = & f^y - f^\gamma 
+ \left(1-\sum_{y^\prime} v_{y^\prime} \right)  
\left[ \sum_x f_x^{'\gamma} \cdot 
	\frac{-\phi_x^y\left(1-\sumt_{y^\prime} v_{y^\prime} \right) +\phi_y^0 
- \sumt_{y^\prime} v_{y^\prime} \phi_x^{y^\prime} }
		{\left(1-\sumt_{y^\prime} v_{y^\prime} \right) ^2} \right] 
\nonumber \\
= & f^y -f^\gamma + \sum_x f^{' \gamma}_x (\phi^\gamma_x - \phi^y_x) 
\nonumber \\
= & \mu_w^y - \mu_w^\gamma = 0 \; , \label{eq:dfter_dvx} 
\end{align}
\end{subequations}
where $\sum_{y^\prime}$ sums over $y^\prime=\alpha,\beta$.
Substituting $x=1,2$ in Eq.~(\ref{eq:dfter_dphiyx}) yields 
Eqs.~(\ref{eq:ternary_equality}a) and (\ref{eq:ternary_equality}b), 
whereas substituting $y=\alpha, \beta$ in Eq.~(\ref{eq:dfter_dvx}) yields 
Eq.~(\ref{eq:ternary_equality}c). Q.E.D.

%In our system of $\chi_{11}=\chi_{22}$, Eqs.~(\ref{eq:f_ternary}) and (\ref{eq:ternary_equality}}) can be simplified by symmetry. Considering that the phase diagram should be symmetric along $\phi_1=\phi_2$, we can suppose $(\phi_1^\beta, \phi_2^\beta)  = (\phi_2^\alpha, \phi_1^\alpha)$ and $\phi_1^\gamma = \phi_2^\gamma$, which yield $v_\alpha=v_\beta$.

Fig.~\ref{fig:3FH_ternary} provides an FH phase diagram with 
both binary and ternary coexistence. In this example,
$\chi_{12}$ is significantly smaller than $\chi_{11}=\chi_{22}$,
resulting in strong effective repulsion between the two sequences. 
Consequently, the two islands of binary coexistence around the 
$\phi_1$- and $\phi_2$-axes intersect with the top-right binary region with
$(\phi_1^\alpha, \phi_2^\alpha) = (\phi_2^\beta, \phi_1^\beta)$,
resulting in a ternary phase separation region corresponding to
the $\alpha\beta\gamma$ triangle and its interior [marked by 
blue lines in Fig.~\ref{fig:3FH_ternary}(a) and turquoise lines 
in Fig.~\ref{fig:3FH_ternary}(b) and (c)]. The thermodynamic stability
of the ternary phase-separated state within this region is illustrated
by the $\Delta f$ quantity plotted in Fig.~\ref{fig:3FH_ternary}(b) and (c);
$\Delta f(\phi_1,\phi_2)$ is the bulk (not-phase-separated) free energy 
$f_{\rm bulk}$ minus the free energy value for the same $\phi_1,\phi_2$
on a plane defined by the three ternary phases (i.e., $\Delta f = 0$ for 
the three points corresponding to the $\alpha$, $\beta$, $\gamma$ phases).
Because $\Delta f > 0$ for 
any other point within the triangular region, the ternary state is more 
stable than the bulk state in this region. 
Moreover, the free energy $f_{\rm sep}$ of any putative binary 
coexistence state of a bulk state within the triangular region must lie 
on a tie line joining two points on the landscape in 
Fig.~\ref{fig:3FH_ternary}(b) and (c). Because $\Delta f>0$
for any point other than the three ternary phases in the entire plotted
region---including points outside the triangular region, $\Delta f > 0$ 
holds also for any putative binary coexistence state for the
bulk state within the triangular region, implying that they are less 
stable than the ternary phase-separated state in the region.

For any given bulk-state $(\phi_1^0,\phi_2^0)$ in the ternary
region, fractional volumes $v_\alpha$, $v_\beta$, 
and $v_\gamma$ in the respective coexisting phases $\alpha$, $\beta$, 
and $\gamma$ are determined by solving
Eq.~\ref{eq:3density} and setting $v_\gamma=1-v_\alpha-v_\beta$.
In terms of the three-dimensional vectors
${\bf \Phi}_0\equiv(\phi_1^0,\phi_2^0,0)$,
${\bf \Phi}_z\equiv(\phi_1^z,\phi_2^z,0)$,
${\bf \Phi}_{0z}\equiv {\bf \Phi}_z-{\bf \Phi}_0$,
and ${\bf \Phi}_{z_1z_2}\equiv {\bf \Phi}_{z_2}-{\bf \Phi}_{z_1}$
where $z,z_1,z_2=\alpha,\beta,\gamma$, 
$v_\alpha=|{\bf \Phi}_{0\gamma}\times {\bf \Phi}_{\beta\gamma}|
/|{\bf \Phi}_{\alpha\gamma}\times {\bf \Phi}_{\beta\gamma}|$,
$v_\beta=|{\bf \Phi}_{0\alpha}\times {\bf \Phi}_{\alpha\gamma}|
/|{\bf \Phi}_{\alpha\gamma}\times {\bf \Phi}_{\beta\gamma}|$
and
$v_\gamma=|{\bf \Phi}_{0\beta}\times {\bf \Phi}_{\alpha\beta}|
/|{\bf \Phi}_{\alpha\gamma}\times {\bf \Phi}_{\beta\gamma}|$.
Because the area of a triangle defined by two vectors is equal to 
half of the magnitude of their cross product, these fractional volumes
correspond to specific ratios of triangular areas
as described in the caption for Fig.~\ref{fig:3FH_ternary}.

\section{Discussion}

\subsection{Insights into cellular binary and ternary IDP phase coexistence}

The present theoretical development bears on the sequence dependence of
multicomponent phase separation in the cell. However, because the
cellular processes involve many species of biomolecules and are extremely
complex~\cite{parker2016,Feric16}, development of treatments much
more elaborated than our simple theories will be needed for quantitative 
comparison with experiments. Nonetheless, it is instructive 
to explore whether our RPA and FH results
are qualitatively consistent with what has been observed experimentally.

Of interest are fibrillarin FIB1 (323 residues) and 
nucleophosmin NPM1 (299 residues) from frog ({\it Xenopus laevis}) 
oocytes. These IDPs tend to demix, exhibiting phase behaviors that 
likely underpin the assembly of nucleolar subcompartments \cite{Feric16}. 
Treating histidine sidechains at {\it p}H $\gtrsim 7$ as neutral,
the net charge of FIB1 is 19 and of NPM1 is $-22$. 
Their charge patterns, as quantified by SCD 
[Eq.~(\ref{eq:SCD})] $=4.126$ and $-0.119$, respectively,
are substantially different. Thus, the tendency for FIB1
and NPM1 to demix is qualitatively in line with the RPA-predicted 
trend in Figs.~\ref{fig:6biCoex} and \ref{fig:SCD_phi_ratios}.

Aqueous solutions with both FIB1 and NPM1 undergo both binary and
ternary liquid-liquid phase separations. In this respect, their experimental
phase diagram in Fig.~4D of Feric et al.~\cite{Feric16} is similar
to our FH phase diagram in Fig.~\ref{fig:3FH_ternary}. The
two regions of binary coexistence of one condensed (around $\alpha$ or
around $\beta$) and one dilute (around $\gamma$) phases in 
Fig.~\ref{fig:3FH_ternary} correspond to their ``FIB1 rich/NPM1 lean'' 
and the ``NPM1 rich/FIB1 lean'' areas,
whereas the ternary coexistence region in Fig.~\ref{fig:3FH_ternary}
with two condensed ($\alpha,\beta$) and one dilute ($\gamma$) phases
[as in Fig.~1(f)] corresponds to their ``3 Phase'' 
area~\cite{Feric16,cliff2017}.

It is noteworthy that our attempts to seek numerical solutions 
to ternary coexistence in the RPA models studied in Fig.~\ref{fig:6biCoex} 
by minimizing Eq.~(\ref{eq:f_ternary}) either ended in failure or 
resulted in solutions with two (among three) phases essentially identical
and thus reduces the solution to that of binary coexistence.
Apparently, ternary coexistence requires an effective intercomponent
repulsion that is substantially stronger, as in Fig.~\ref{fig:3FH_ternary}, 
than that posited by RPA. The reason is that RPA constrains
the intercomponent interaction strength $\chi_{12}$ to approximately 
the geometric mean of the two intracomponent interaction strengths
$\chi_{11}$ and $\chi_{22}$ (Sec.~\ref{sec:imitate}) and therefore 
$\chi_{12}\ll \chi_{11},\chi_{22}$ is highly unlikely if not impossible.

This consideration suggests that difference in charge pattern alone may 
be insufficient to account for the rather strong effective repulsion
between FIB1 and NPM1, although the impact of them being 
not very close to being neutral remains to be investigated.
(Unlike the model KE sequences in Fig.~\ref{fig:6biCoex} with zero net charge
or Ddx4 with a charge ratio $=$ (net charge/chain length) 
$=-1.7\%$~\cite{Lin16},
their charged ratios are, respectively, $+5.9\%$ and $-7.4\%$).
In addition to the electrostatic interactions among FIB1 and NPM1, other 
driving forces surely also contribute to their phase behaviors. 
For example, the presence of ribosomal RNA (rRNA) appears to be important;
and the role of rRNA and inter-phase surface tensions have been modeled 
in lattice simulations to rationalize the
FIB1/NPM1 droplet-in-droplet organization~\cite{Feric16}.
Building on our findings, much effort will be required to 
ascertain the precise role played by charge pattern mismatch in this 
intriguing phenomenon.

\subsection{Cooperativity driven by concentration-dependent relative 
permittivity}

Most analytical formulations for charged polymer solutions, including common 
RPA theories, treat the relative permittivity $\epsilon_{\rm r}$ of the 
solution as a constant independent of polymer concentration.  However, as we 
noted recently~\cite{Lin17a}, because of the significantly different 
permittivities of water ($\ew \approx 80$) and protein 
($\ep \approx 2$--4)~\cite{Leach}, the effective $\epsilon_{\rm r}$ of
a protein solution can change dramatically with protein concentration. 
Indeed, protein-dependent variations of dielectric properties of the 
aqueous medium have been shown to be relevant to globular protein stability
in thermophilic species \cite{Elcock98,HX_zhou02,Charlie04,Kings17}. 
Because of the anticipated importance of dielectric properties to IDP 
energetics such as enabling a greatly enhanced propensity to phase 
separate~\cite{Lin17a}, here we expand our theoretical exploration 
to two-IDP systems and consider in more detail the physical basis
of various effective medium approximations that may be applied to
estimate $\epsilon_{\rm r}$ of IDP solutions. Since the scope of 
this exploration is limited to establishing certain general
principles, for simplicity of the discussion we let $\ep$
be the relative permittivity of both IDP species in the system such that 
$\epsilon_{\rm r}(\phi_1,\phi_2)=\epsilon_{\rm r}(\phi)$ where
$\phi=\phi_1+\phi_2$.

Previously we considered two models for $\epsilon_{\rm r}$~\cite{Lin17a},
namely the ``slab model'' derived by considering a planar electric
capacitor, wherein
\begin{equation}
\epsilon_{\rm r}^{\rm Slab}(\phi) = \frac{\ep\ew}{\phi\ew + (1-\phi)\ep} \; ,
	\label{eq:eps_r_slab}
\end{equation}
which corresponds to Eq.~(3) of Bragg and Pippard when the depolarizing 
coefficient $L=0$~\cite{bragg1953}, and the Clausius-Mossotti (CM) model
predicting
\begin{equation}
\epsilon_{\rm r}^{\rm CM}(\phi) = \frac{1+2[(1-\phi)\gw + 
\phi \gp ]}{1-[(1-\phi)\gw + \phi \gp ]} \; ,
	\label{eq:eps_r_CM}
\end{equation}
where
\begin{equation}
\gw \equiv \frac {\ew -1}{\ew + 2} \; ,
\quad \quad
\gp \equiv \frac {\ep -1}{\ep + 2} \; 
	\label{eq:gamma_CM}
\end{equation}
are proportional to the CM expression for molecular 
polarizability~\cite{CM_equation}. The $\phi$-dependences of 
$\epsilon_{\rm r}^{\rm Slab}(\phi)$ and $\epsilon_{\rm r}^{\rm CM}(\phi)$
are very similar (Fig.~11 of \cite{Lin17a}). Both models recognize
amino acid residues and water as materials with excluded 
volume such that the contributions of their molecular polarizabilities to the
overall dielectric property of the medium are against a vacuum background.

\begin{figure}%[htbp] 
\centering
\includegraphics[width=0.8\columnwidth]{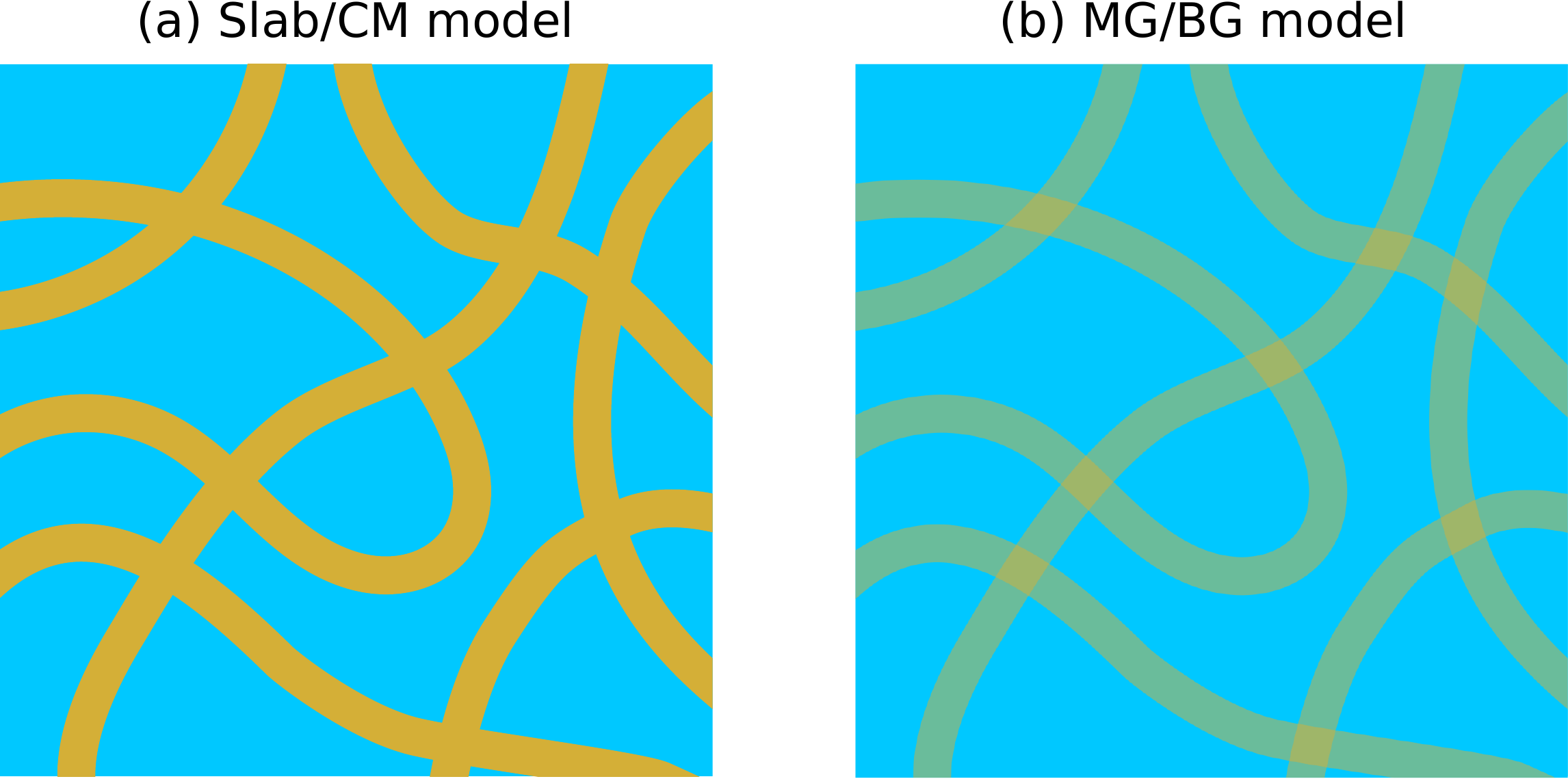} 
\caption{Physical pictures underlying different 
effective medium approximations for the relative 
permittivity of aqueous IDP solutions. IDP chains are golden, water is 
depicted as the blue background. (a) The slab and CM models
assume that water is excluded from the volume occupied by the IDPs
and therefore does not contribute to the dielectric effect inside
IDP volume. (b) In contrast, the MG and BG models view 
the IDP chains as embedding in an all-permeating water background, 
such that water contributes to the dielectric effect both outside and 
inside of the IDP volume. Here this assumption is indicated 
schematically by the translucency of the IDP chains.
}
   \label{fig:permittivity_model}
\end{figure}

An alternate approach to effective medium is the Maxwell Garnett (MG) 
model~\cite{Markel16} that pictures the effective dielectric property
of a composite material as arising from embedding a component (IDP
in our case) into an all-permeating background medium (water in our case).
IDP excluded volume is not taken into account in this approach.
Applying this method to our IDP solution yields
\begin{equation}
\epsilon_r^{\rm MG}(\phi) = 
\ew \frac{1+2\phi\gamma_{\rm MG}}{1-\phi\gamma_{\rm MG}},
\end{equation}
where
\begin{equation}
\gamma_{\rm MG} \equiv \frac {\ep -\ew}{\ep +2\ew} \; .
	\label{eq:gamma_MG}
\end{equation}
Comparing the $\gamma_{\rm p}$ expression in Eq.~(\ref{eq:gamma_CM}) with 
Eq.~(\ref{eq:gamma_MG}) indicates
that $\gamma_{\rm MG}$ corresponds---up to a constant---to an 
{\it effective} molecular polarizability
of IDP material, not against vacuum but rather in a water background 
($\gamma_{\rm p}$ is mathematically equal to $\gamma_{\rm MG}$ when $\ew\to 1$,
thus MG is related to CM in this respect~\cite{Tsukerman2017}).
Another approach known as the Bruggeman (BG) model~\cite{Markel16}
is an $\ep\leftrightarrow\ew$ symmetrized form of MG. The BG $\phi$-dependent
permittivity is given by
\begin{equation}	
\epsilon_r^{\rm BG}(\phi) \equiv \frac{b_{\rm BG}(\phi) + 
\sqrt{8\ew\ep+b_{\rm BG}(\phi)^2}}{4} \; ,
\end{equation}
where
\begin{equation}
b_{\rm BG}(\phi) = [2\phi-(1-\phi)]\ep + [2(1-\phi)-\phi]\ew \; .
\end{equation}
A graphical illustration of the different physical pictures assumed by 
the slab/CM versus the MG/BG approaches is provided by
Fig.~\ref{fig:permittivity_model}.
Leaving aside questions about the appropriateness of 
their respective physical pictures for the moment, we first 
examine the properties of these $\epsilon_{\rm r}(\phi)$'s and their
implications for IDP phase separation.

\begin{figure}%[htbp] 
%\centering
\includegraphics[width=\columnwidth]{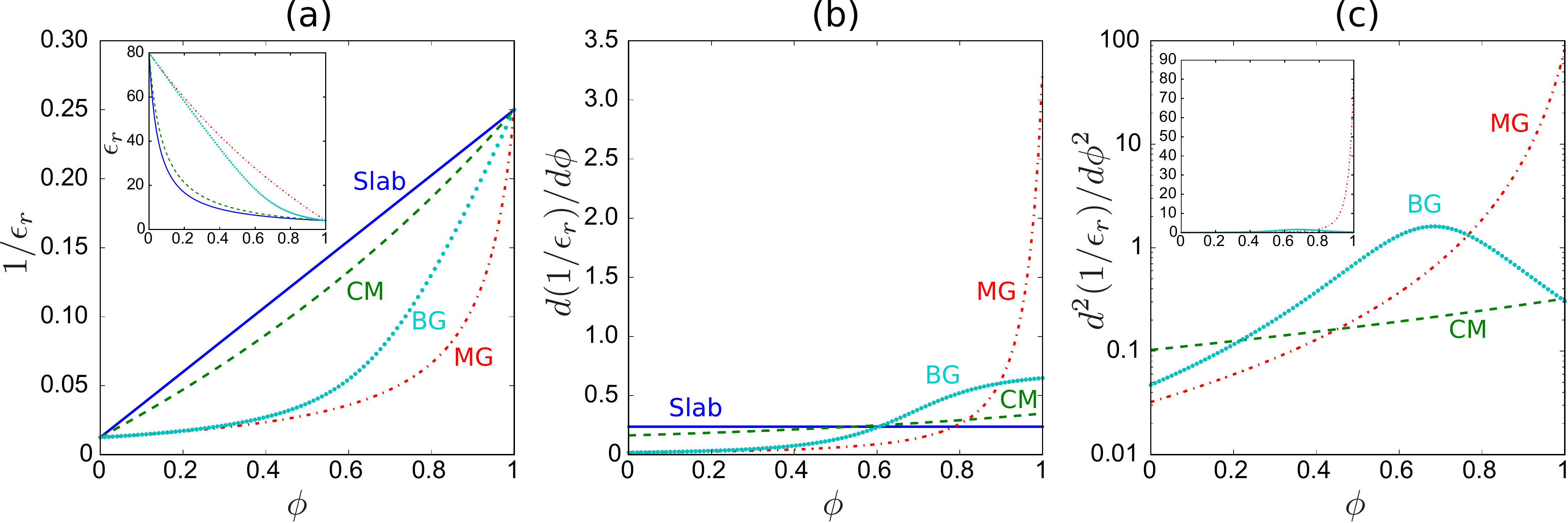} 
\caption{IDP-concentration-dependent relative permittivity estimated
by different effective medium approximations. 
Plotted data are for $\epsilon_{\rm w}=80$, $\epsilon_{\rm p}=4$.
(a) $1/\epsilon_r(\phi)$ of the four models: Slab, CM, MG, and BG. 
Inset shows $\epsilon_r(\phi)$. 
(b) First order derivatives of $1/\epsilon_r(\phi)$ of the four models. 
(c) Second order derivative of $1/\epsilon_r(\phi)$ of the 
CM, MG, and BG models are plotted in logarithmic scale. Data for the
slab model is not shown because its second-order derivative vanishes.
Inset shows the same data in linear scale. 
}
   \label{fig:permittivity_trends}
\end{figure}

Variations of several properties of the $\epsilon_{\rm r}(\phi)$'s 
are shown in Fig.~\ref{fig:permittivity_trends}.  Although all four models 
give $\epsilon_{\rm r}(\phi=0)=\ew$ and $\epsilon_{\rm r}(\phi=1)=\ep$ as 
they should, the $\phi$-dependences of the slab/CM models are very
different from that of the MG/BG models. Here we are more
interested in $1/\epsilon_{\rm r}$ than $\epsilon_{\rm r}$ itself
because $1/\epsilon_{\rm r}$ is directly proportional to Coulomb energy. 
Fig.~\ref{fig:permittivity_trends}(a) shows that $1/\epsilon_{\rm r}$ 
of the slab model is linear in $\phi$, that of CM is approximately linear;
but the $1/\epsilon_{\rm r}$'s of MG and BG increase very little
when $\phi$ is small and exhibit rapid increases toward the $1/\ep$
value for $\phi=1$ only for $\phi \gtrsim 0.8$ and $0.6$, respectively. 
The linear and near-linear $\phi$-dependences of the $1/\epsilon_{\rm r}$'s 
for the slab and CM models are underscored by their small 
first and second derivatives in $\phi$, whereas the sharp rises of 
$1/\epsilon_{\rm r}$ near $\phi\approx 1$ for the MG and BG models 
are illustrated by their large $\phi$-derivatives for $\phi \gtrsim 0.7$
[Fig.~\ref{fig:permittivity_trends}(b) and (c)].

\begin{figure}%[htbp] 
\centering
\includegraphics[width=\columnwidth]{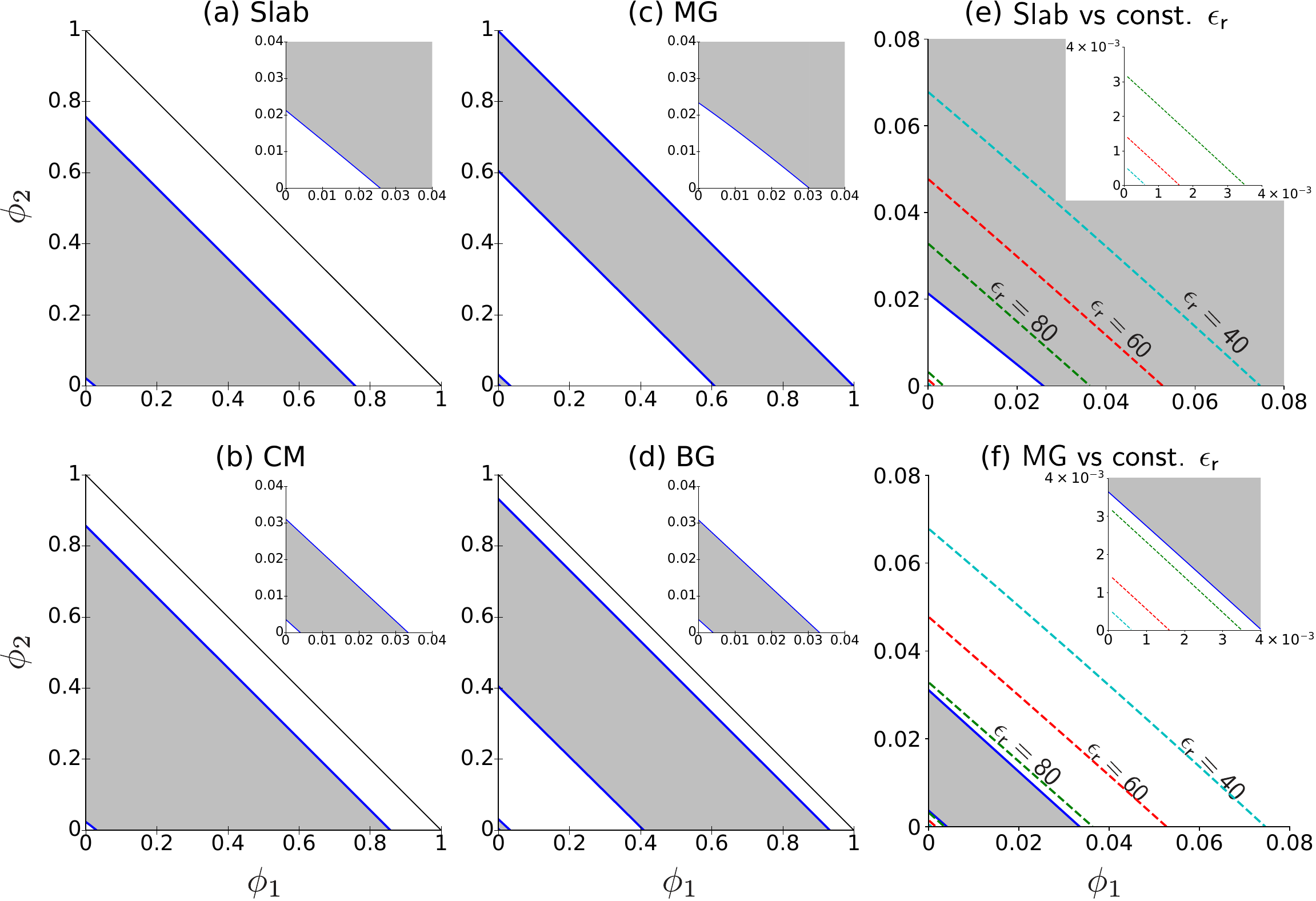} 
\caption{RPA-predicted spinodal instability of a two-sequence aqueous solution 
with an IDP-concentration-independent or an IDP-concentration-dependent
relative permittivity. Results in (a)--(d) are for the (seq1, seq2) system
at $T_0^* = 0.05$ computed using the four different models of 
$\epsilon_{\rm r}(\phi)$ described in the text, with $\ew =80$, $\ep = 4$, 
and $\phi=\phi_1+\phi_2$. Spinodal regions are shaded and bounded by solid 
blue lines. The inclined solid black line $\phi_1+\phi_2=1$ is the 
volume-conservation
boundary. (a) Slab model, exhibiting only a single spinodal 
region; (b) Clausius-Mossotti model, also showing a single spinodal region; 
(c) Maxwell Garnett model, two spinodal regions are observed; note that
the upper spinodal boundary is very close to but not identical with 
$\phi_1+\phi_2=1$. (d) Bruggeman model, showing also two spinodal 
regions. 
(e) Comparing the dilute spinodal boundary in the
slab model (blue solid line) against the spinodal boundaries for
RPA models with $\phi$-independent permittivity $\epsilon_{\rm r} = 40$, $60$, 
and $80$ (dashed lines) at the same temperature $T_0^*$.
(f) Similar to (e) but here we consider the MG model instead of the slab model. 
In (e) and (f), all dash lines higher than the
blue solid line are condensed-phase boundaries for the
constant-$\epsilon_{\rm r}$ models. 
Insets in (a)--(f) are zoom-in's that provide a closer look at the dilute
spindoal phase boundaries.
}
   \label{fig:sp_dic_permittivity}
\end{figure}

As described in our previous work~\cite{Lin17a}, when permittivity 
becomes a function of IDP concentration, the last subtraction of 
${\cal G}(k)$ in Eq.~(\ref{eq:fel_integral}) has to be modified
because ${\cal G}(k)$ is no longer linear in $\phi$'s~\cite{Lin17a}. 
A straightforward generalization of Eqs.~(68) and (69) of Ref.~\cite{Lin17a}
to the present case with two IDP sequences (but now with neither salt
nor counterions) leads to the following replacement for the RPA expression 
in Eq.~(\ref{eq:fel_integral}) to accommodate a $\phi$-dependent 
$\epsilon_{\rm r}$:
\begin{equation}
f_{\rm el}(\phi_1,\phi_2)  = \int_0^\infty \frac{d k k^2}{4\pi^2} 
\left\{ \ln \left[ 1+ {\cal G}_1(k) \right] - {\cal G}_2 (k) \right\} \; ,
\end{equation}
where
\begin{equation}
\begin{aligned}
{\cal G}_1(k) = & \frac{4\pi}{k^2(1+k^2)T^*_0\epsilon_r(\phi)}
\left[ 
\langle \sigma_1 | \hat{G}_{11}(k) | \sigma_1 \rangle  + 
\langle \sigma_2 | \hat{G}_{22}(k) | \sigma_2 \rangle 
\right], \\
{\cal G}_2(k) = & \frac{4\pi}{k^2(1+k^2)T^*_0\epsilon_r(\phi)} 
\left[ 
\frac{\phi_1}{N_1}\sum_i\left|\,\sigma_1^{(i)} \,\right| 
	+ \frac{\phi_2}{N_2}\sum_i\left|\,\sigma_2^{(i)}\,\right|\,
\right] \; , 
\end{aligned}
\end{equation}
and, following Eq.~(67) of Ref.~\cite{Lin17a},
\begin{equation}
T_0^* \equiv 4\pi\epsilon_0 k_{\rm B}Tb/e^2 = T^*/\epsilon_{\rm r} \; .
\end{equation}
The resulting spinodal phase diagrams predicted by the four 
$\epsilon_{\rm r}(\phi)$ models are shown in 
Fig.~\ref{fig:sp_dic_permittivity} for 
the (seq1, seq2) pair at $T_0^* = 0.05$. This temperature is chosen
to facilitate comparison with constant-$\epsilon_{\rm r}$ results
because $T_0^* = 0.05$ corresponds to $T^*= 4$ when $\ew =\ep=80$.

Fig.~\ref{fig:sp_dic_permittivity}(a)--(d) shows that all four 
$\epsilon_{\rm r}(\phi)$ models have large spinodal regions
extending to $\phi_1+\phi_2\approx 0.8$.
These spinodal regions are much larger than those predicted
by constant-$\epsilon_{\rm r}$ theories
[Fig.~\ref{fig:sp_dic_permittivity}(e), (f)], pointing once again to 
a significant cooperative effect arising from a decrease in
permittivity upon IDP condensation which in turn increases
electrostatic attraction and hence more IDP condensation~\cite{Lin17a}.
For instance, the condensed-phase volume fractions of the slab and CM 
models $\approx 0.8$, which represents a $>20$-fold increase 
from the condensed-phase volume fraction of $\lesssim 0.03$ for a constant 
$\epsilon_{\rm r}=\ew =80$ [Fig.~\ref{fig:sp_dic_permittivity}(e)].
The corresponding enhancement in the MG and BG models are even
more prominent---their condensed-phase volume fraction almost
reaching the $\phi_1+\phi_2=1$ limit [Fig.~\ref{fig:sp_dic_permittivity}(c),
(d)]---because of the sharp rises of their $\epsilon_{\rm r}(\phi)$'s 
when $\phi\to 1$ (Fig.~\ref{fig:permittivity_trends}). 
Although the precise quantitative impact of $\epsilon_{\rm r}(\phi)$
remains to be ascertained experimentally, our theoretical
results suggest strongly that $\phi$-dependent relative permittivity
should play a significant role in IDP phase separation, and
that such a physical cooperative effect may help rectify some of the
RPA-predicted condensed-phase volume fractions 
[e.g. those in Fig.~\ref{fig:6biCoex} (a)--(d)] that seem 
unrealistically low.

Interestingly, while the slab and CM models enlarge a single spinodal region
vis-\`a-vis that for constant-$\epsilon_{\rm r}$,
the MG and BG models produce an additional spinodal region
close to the $\phi$-$\phi_2$ origin. This region is similar in scope to 
the constant-$\epsilon_{\rm r}$ spinodal region, and is
well separated from the
MG and BG models' lower boundaries at $\phi_1+\phi_2 \approx 0.6$ (MG) 
or 0.4 (BG) for their main (much larger) spinodal regions
[Fig.~\ref{fig:sp_dic_permittivity}(c)--(f)].
This feature likely arises from the fact that 
$1/\epsilon_{\rm r}$'s for MG and BG barely change
for $\phi \lesssim 0.2$ [Fig.~\ref{fig:permittivity_trends}(a)],
thus the behaviors of these models at small $\phi$'s should be similar
to those of constant-$\epsilon_{\rm r}$ models.
For larger $\phi$'s, however, because of the rapid increase
of their $1/\epsilon_{\rm r}$'s with $\phi$, 
the MG and BG models become similar to the slab and CM models.

\begin{figure}%[htbp] 
\centering
\includegraphics[width=0.75\columnwidth]{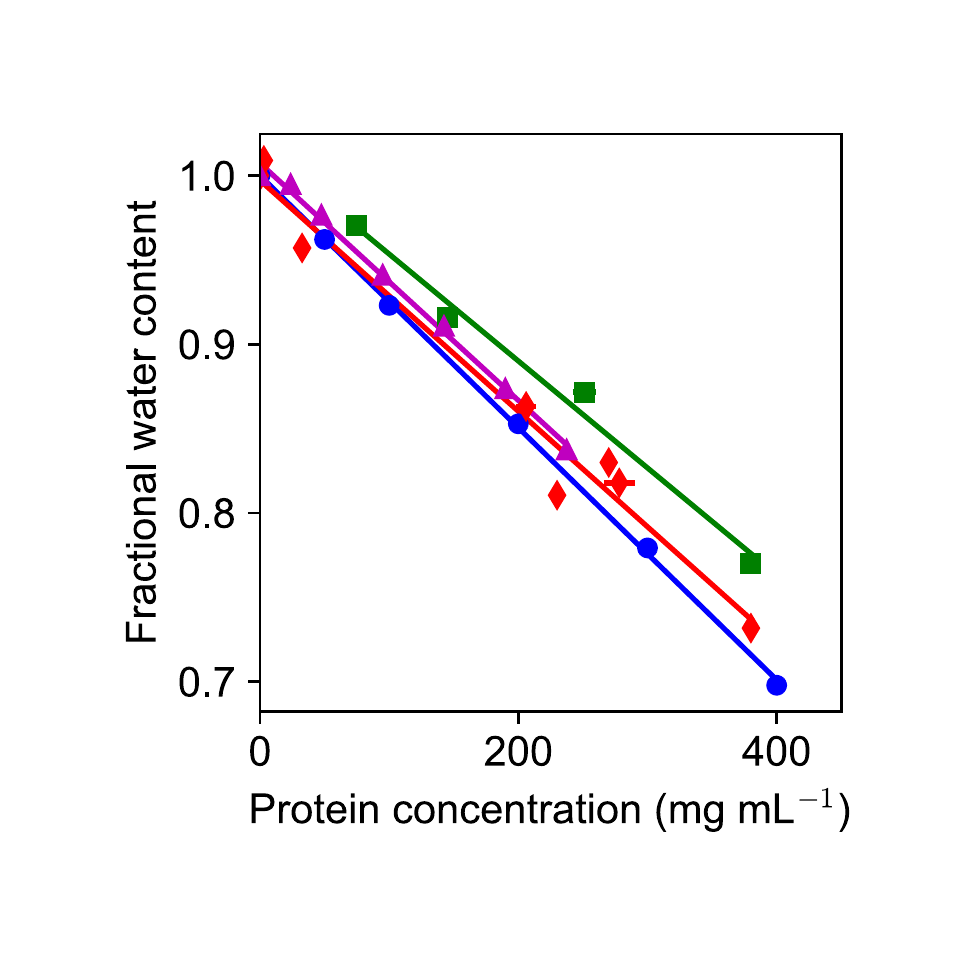} 
\vskip -1.0cm
\caption{Fractional water content of protein solution samples measured 
by NMR (see Sec.~\ref{sec:methods}): BSA (blue, circles), HEWL 
(magenta, triangles), \ftoa (green, squares), \cond (red, diamonds).
The slopes of the fitted lines give the densities of these proteins in 
water as $-1/({\rm slope})$, yielding densities (in units of
g cm$^{-3}$) of $1.337\pm 0.015$ (BSA), $1.432\pm 0.041$ (HEWL),
$1.465\pm 0.099$ (\cond), and $1.603\pm 0.139$ (\ftoa).
}
   \label{fig:volume}
\end{figure}

Although the slab/CM and MG/BG models produce similarly expanded
spinodal regions for the example in Fig.~\ref{fig:sp_dic_permittivity},
the difference in their predicted $\epsilon_{\rm r}(\phi)$'s do
have important implications on the energetics of IDP phase separation.
For example, the condensed-phase volume fraction of Ddx4 was
recently determined to be $\approx 0.15$--$0.2$ \cite{Brady:2017}.
At $\phi\approx 0.2$, the slab/CM models posit a significant cooperative
effect in favor of phase separation, but the MG/BG models suggest
that such cooperativity is negligible unless $\phi\gtrsim 0.6$
(Fig.~\ref{fig:permittivity_trends}). Issues related to effective
medium approximations can be intricate, as witnessed by the extensive 
materials-science literature on the topic \cite{Markel16,Tsukerman2017}.
Nevertheless, for IDP solutions, our intuition is that the slab/CM scenario 
is more physically plausible than the MG/BG scenario.  Consistent with the 
slab/CM picture in Fig.~\ref{fig:permittivity_model}(a), just like
dissolved folded proteins, dissolved IDPs occupy volumes excluded to water. 
Dissolved IDPs and folded proteins have similar densities 
(Fig.~\ref{fig:volume}) of
$1.32$ -- $1.52$ g cm$^{-3}$ \cite{protein_density2004} and hence
similar partial molar volumes on a per-gram basis.
In contrast, the MG/BG picture in Fig.~\ref{fig:permittivity_model}(b) 
invokes a negative effective molecular polarizability 
for IDP that counteracts an all-permeating water medium 
($\gamma_{\rm MG}<0$ because $\ep<\ew$). This scenario is apparently
at odds with the reality of IDP excluded volume, suggesting that
while the MG/BG models may be applicable in certain
solid-state situations \cite{Tsukerman2017}, they may be problematic
for IDP solutions. A definitive resolution of this question 
awaits further theoretical and experimental investigation. 

\section{Conclusions}

To recapitulate, we have taken a step to address the sequence-phase
relationship in regard to how mixing/demixing of IDP components
in membraneless organelles are governed by the
IDPs' amino acid sequences. Going beyond mean-field FH and OV approaches,
RPA provides an approximate physical
account of sequence effects~\cite{Lin16,Lin17a,Lin17b}.
Our findings point to a multivalent, stochastic, ``fuzzy''
mode of molecular recognition in that mixing the populations of a pair of IDP
sequences is favorable if their charge patterns are similar whereas
population demixing is promoted when their charge patterns are very different.
For the examples studied, a quantitative correlation is observed between 
the RPA-predicted tendency for a pair of sequences to demix in two (binary) 
coexisting phases and the difference in their 
SCD parameters. This predicted trend is qualitatively in line with the 
observed demixing of the nucleolar IDPs FIB1 and NPM1 because they have very 
different SCD's, although a comparison of the experimental FIB1/NPM1 
phase diagram~\cite{Feric16} with our RPA and FH results indicates that 
inclusion of non-electrostatic interactions as well as more biomolecular 
species in the analysis will be necessary for a quantitative theoretical 
account of sequence-specific ternary coexistence. It should also be noted
that our current RPA formulation does not consider counterion condensation, 
which can be important for IDP sequences with high net charges.
A recent transfer matrix theory \cite{Sing17} should be helpful in tackling
such situations, although as it stands this theory does not address 
sequence dependence. Despite our theory's limitations,
the simple principles of sequence dependence suggested by 
the present effort should already be testable by in vitro experiments 
on IDP polymers~\cite{chilkoti2015}.
As illustrated by our consideration of IDP-concentration-dependent 
permittivity, theoretical study of IDP phase separation is only
in its infancy. The logical next steps in the development of RPA
theory include extending to systems with more than two sequences
and sequences with non-zero net charges. Much biophysics of IDP phase 
separation remains to be discovered.
\\

\section*{Acknowledgements}
We thank Alaji Bah, Robert Vernon (Hospital for Sick Children), 
Lewis Kay (University of Toronto), Che-Ting Chan (Hong Kong University 
of Science and Technology), Kingshuk Ghosh (University of Denver), 
Pak-Ming Hui (Chinese University of Hong Kong), and Huan-Xiang Zhou
(University of Illinois at Chicago) for helpful 
discussions. We are also grateful to an anonymous referee for
suggestions that led to the pedagogical graphics in Fig.~1. 
This work was supported by Canadian Cancer Society Research 
Institute grant no. 703477, Canadian Institutes of Health Research grant 
MOP-84281, and computational resources provided by SciNet of 
Compute/Calcul Canada.
\\

\section*{References}
 
\bibliographystyle{iopnum}
\bibliography{2proteins_v1}

\end{document}